\documentclass[reprint, superscriptaddress, amsmath, amssymb,pra]{revtex4-1}
\usepackage{graphicx}
\usepackage{bbold}
\usepackage{textcomp}
\usepackage{gensymb}
\usepackage{xcolor}
\usepackage[normalem]{ulem}
\usepackage{lineno}
\setcitestyle{super}
\usepackage{color}
\usepackage{makecell}
\usepackage{mathtools}

\usepackage{soul}

\usepackage{hyperref} 

\graphicspath{{./figs/},{./}}
\graphicspath{{./figs/},{./}}
\newcommand{\SAF}{\textrm{SAF}}

\newcommand{\AIR}{\textrm{AIR}}

\newcommand{\dTmax}{\ensuremath{\Delta T_\mathrm{max}}}
\newcommand{\Tmax}{\ensuremath{T_\mathrm{max}}}

\begin{document}

\title{Universalities of Asymmetric Transport in Nonlinear Wave Chaotic Systems}

\author{Cheng-Zhen Wang}
\affiliation{Wave Transport in Complex Systems Lab, Department of Physics, Wesleyan University, Middletown, CT-06459, USA}
\author{Rodion Kononchuk}
\affiliation{Wave Transport in Complex Systems Lab, Department of Physics, Wesleyan University, Middletown, CT-06459, USA}
\author{Ulrich Kuhl}
\affiliation{Université Côte d’Azur, CNRS, Institut de Physique de Nice (INPHYNI), 06108, Nice, France}
\author{Tsampikos Kottos}
\affiliation{Wave Transport in Complex Systems Lab, Department of Physics, Wesleyan University, Middletown, CT-06459, USA}


\begin{abstract}
The intrinsic dynamical complexity of classically chaotic systems enforces a universal description of the transport properties of their
wave-mechanical analogues. These universal rules have been established within the framework of linear wave transport, where non-
linear interactions are omitted, and are described using Random Matrix Theory (RMT). Here, using a nonlinear complex network of
coaxial cables (graphs), we exploit both experimentally and theoretically the interplay of nonlinear interactions and wave chaos. We
develop general theories that describe our asymmetric transport (AT) measurements, its universal bound, and its statistical description
via RMT. These are controlled by the structural asymmetry factor (\SAF{}) characterizing the structure of the graph. The \SAF{}
dictates the asymmetric intensity range (\AIR{}) where AT is strongly present. Contrary to the conventional wisdom that expects losses
to deteriorate the transmittance, we identify (necessary) conditions for which the \AIR{} (AT) increases without deteriorating the AT
(\AIR{}). Our research initiates the quest for universalities in wave transport of nonlinear chaotic systems and has potential applications
for the design of magnetic-free isolators.
\end{abstract}

\maketitle

{\bf Introduction -} Wave chaos is an interdisciplinary field of physics that aims to describe the properties of wave systems with
underlying classical chaotic dynamics. At its foundations, is the assumption that the generated classical complexity enforces a universal
wave description that trespass physical frameworks ranging from atomic nuclei, optical and microwave mesoscopic systems, to macroscopic
acoustic and even ocean waves. These universal laws can be described by phenomenological mathematical theories like Random Matrix
Theory (RMT). Despite the success of these methodologies, still, their validity is confined by the assumption that wave-matter nonlinear
interactions are not present -- a condition that allows us to utilize the superposition principle and scale invariance. At the same time,
nonlinear mechanisms are abundant in nature and, in many cases, offer exciting new opportunities to manipulate waves and develop
novel structures with novel functionalities. In this respect, one can only imagine the range of new opportunities that will become available
with the development of a predictive wave transport framework which allows for the coexistence of chaos and nonlinearity.

\begin{figure*}
\centering
\includegraphics[width=0.75\linewidth]{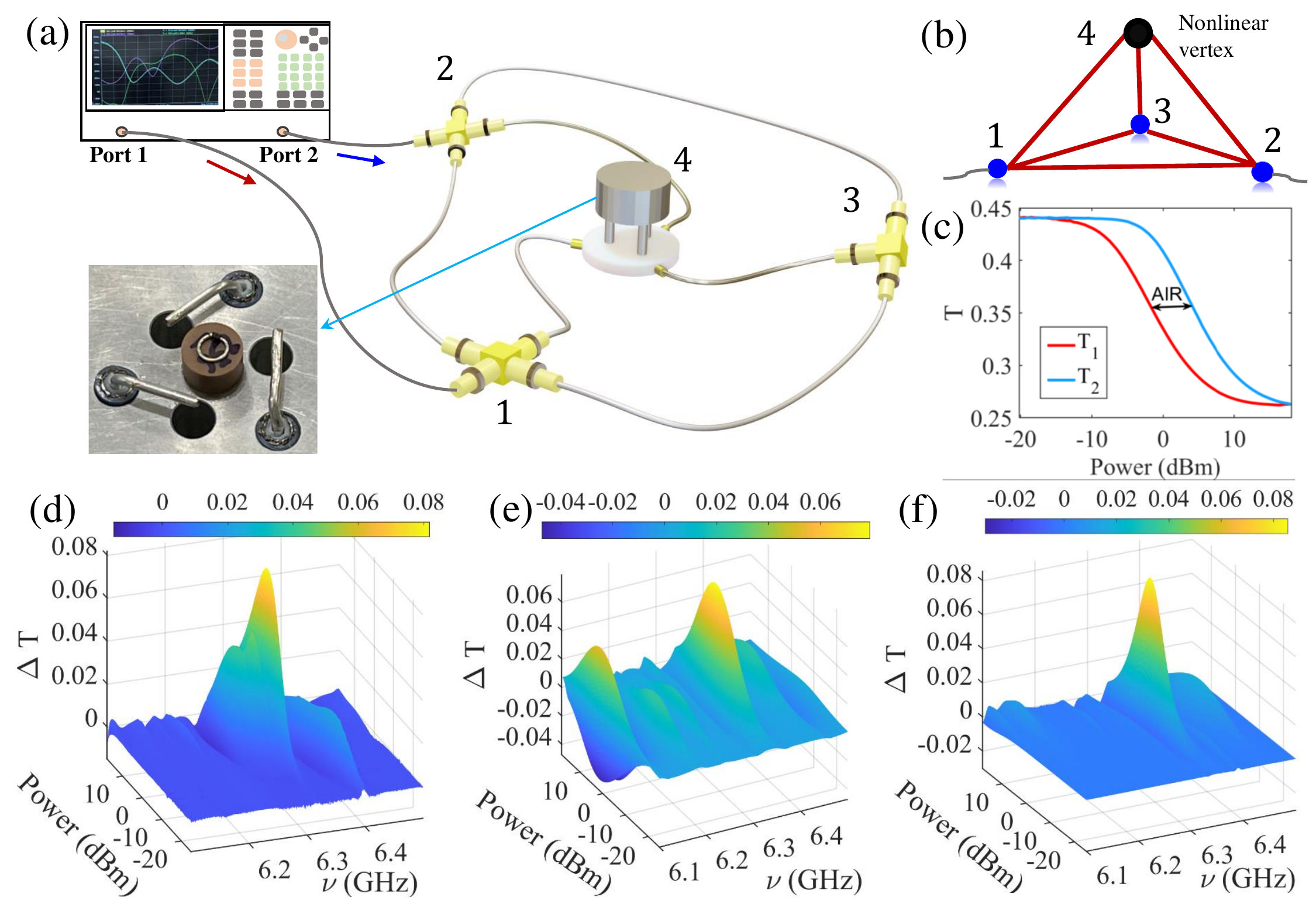}
\caption{ \label{fig1}
{\bf Experimental setup of a nonlinear microwave graph and transmission asymmetry.} (a) A microwave tetrahedron graph. The coaxial
cables are connected by T or double T-junctions at each of the vertices $n=1,2$, and $3$. Vertex $N=4$ consists of three kink antennas
coupling to a cylindrical dielectric resonator that is inductively coupled to a ring antenna which is short-circuited with a nonlinear diode
(see left inset). (b) A schematics of the tetrahedron graph shown in (a). (c) Measured transmittance $T_{1}$ (from port 1 to port 2) and
$T_{2}$ (from port 2 to port 1) at fixed frequency ($\nu=6.327$\,GHz) as a function of the input power showing asymmetric transport.
(d)-(f) The transmittance difference $\Delta T\equiv T_{2}-T_{1}$ for an incident wave (of same amplitude and frequency) as a function
of frequency and input power. (d) Experimental data; (e) simulations; and (f) using a resonant-graph modeling.
}
\end{figure*}

Here we make a first effort in the direction of creating a theory of nonlinear wave chaos. We focus our attention to the technologically relevant
question of asymmetric transport (AT) and its statistics. Asymmetric/nonreciprocal devices such as isolators and circulators are routinely used
in wireless and optical communications, radar and LiDAR technologies and integrated photonic circuits at microwave and optical frequencies  \cite{caloz2018electromagnetic}. Their principle of operation relies in the violation of reciprocity which is typically achieved (in linear structures)
by using an odd-vector bias (e.\,g.\ an external magnetic field)~\cite{pozar2011microwave,potton2004reciprocity} or by violating the time-
invariance via a spatio-temporal modulation~\cite{sounas2017non}. Utilizing nonlinearities for the realization of AT is an alternative promising
approach.Unfortunately, most of the existing studies, either analyze AT in simple nonlinear systems~\cite{wang2013fundamental,sounas2017time,
sounas2018fundamental,sounas2018broadband,cotrufo2021nonlinearity2}, or they address the coexistence of chaos and nonlinearities
\cite{ZOAA17,ZOAA19,Sven11,Sven16} without paying attention to AT and its statistical description.

Here, we describe asymmetric transport (AT) via universal rules imposed by the underlying classical chaotic system. Our analysis utilizes a
prototype platform for wave chaos, i.e., complex networks of coaxial cables (graphs), see Figs.~\ref{fig1}a-b. The motivation for this choice
is twofold: (a) graphs have been established as a friendly system where both RMT~\cite{Casati1980,Bohigas1984} and semiclassical tools
\cite{Haake2018} can be deployed successfully in describing transport~\cite{kottos1997quantum,kottos1999periodic,kottos2000chaotic,
texier2001scattering,kottos2003quantum,Schanz2003,dietz2017nonuniversality, chen2020perfect,chen2021statistics,Gnutzmann2004,
Gnutzmann2006,Gnutzmann2008,Pluhar2013,Pluhar2014,Haake2018}; (b) the system is experimentally accessible in a variety of wavelengths
-- from acoustics to microwaves and optics. Our analysis highlights an intimate relation between the AT properties of a nonlinear chaotic system
and the structural asymmetry factor (\SAF{}) that is determined by the structural complexity of the underlying linear structure. We find that
\SAF{} dictates the asymmetric intensity range (\AIR{}) defined as the ratio of input powers injected from opposite directions which lead
to the same transmittance (see Fig.~\ref{fig1}c). Furthermore, we have derived a general expression for the upper bound of the AT in terms
of losses and other system-specific characteristics. The case of lossless graphs reproduces previously established bounds~\cite{sounas2018fundamental,
cotrufo2021nonlinearity1} and it is recovered as a limit of this general expression. Using these results, we have identify necessary conditions
for a class of lossy non-linear chaotic scattering settings whose transmission asymmetry bound exceeds the one of the corresponding lossless
analogues. We demonstrate experimentally, that this class does not degrade the transmission asymmetry at all -- instead it enhances the \AIR{}.
The generality of our results are established using an RMT that incorporates nonlinearities and show theoretically and experimentally that the
distribution of the rescaled transmission asymmetries, i.\,e., $\widetilde {\Delta T}\equiv \Delta T/\dTmax$ is nicely reproduced by this theory.

\section*{Results}
\noindent{\bf Experimental Implementation --}
A nonlinear microwave graph consists of coaxial cables (Huber+Suhner S 04272) connected together via $n=1,\cdots, N$ junctions (vertices).
The electrical permitivity of the cables was found to be $\epsilon\approx 1.56(\pm0.07)+i0.0015(\pm0.0005)$ indicating the presence of uniform
losses (see Methods). The number of coaxial cables (bonds) emanating from a vertex $n$ is its valency $v_n$ and the total number of directed
bonds (i.e., discerning $B\equiv n\rightarrow m$ and $\bar B\equiv m\rightarrow n$) is $2V= \sum_{n=1}^N v_n$. For the tetrahedron graph
shown in Figs.~\ref{fig1}a,b the vertices are Tee-junctions and $N=4, v_n=3$ ($n=1,\cdots,4$). The local nonlinearity is always incorporated
at the $N$-th vertex. It is implemented via a dielectric resonator coupled inductively to a diode from the top, and to three coaxial cables that
form a Tee-junction-like vertex (see Methods and SM Sec. \ref{Appendix-Para-Fitting}).

Figure~\ref{fig1}c shows the measured transmissions $T_{1}$ (port 1 to 2) and $T_{2}$ (port 2 to 1) for a fixed frequency $\nu=6.327$\,GHz
as a function of the input power. We find that the non-linearity is operational as a strong nonlinear dependence of the transmissions on the input
power is observed. Additionally, one can extract the maximal transmission \dTmax{} difference as well as the \AIR{}. In Fig.~\ref{fig1}d we
show the measured transmission difference $\Delta T = T_{2}-T_{1}$ as a function of the input power and frequency $\nu$.

\noindent{\bf Theoretical modeling -- }
The theoretical analysis utilizes a standard open quantum graph description \cite{kottos2000chaotic} with the modification that the $N$-th vertex
is now nonlinear (details are presented in the Methods). For a compact description of the nonlinear scattering process it is useful to introduce the
scattering vector field $\Phi^{(\alpha)} = (\phi_1^{(\alpha)},\phi_2^{(\alpha)},\cdots, \phi_{N}^{(\alpha)})^T$ where $\phi_{n}^{(\alpha)}$
indicates the field amplitude associated with the vertex $n$ while the superindex $\alpha=1,2$ indicates
the incident TL. The scattering vector field $\Phi^{(\alpha)}$ satisfies the matrix equation (see SM Sec. \ref{Appendix-GGF})
\begin{align} \label{GMM}
	(M + M_{NL} + iW^TW)\Phi^{(\alpha)} = 2iW^TI^{(\alpha)}\,,
\end{align}
where the two-dimensional vector $I^{(\alpha)}$ with components $I^{(\alpha)}_{\mu} = A_{\mu}\delta_{\alpha,\mu}$ describes the
amplitude of the incident field of the channel $\alpha$ that has been used to inject the wave, and $W$ is a $2\times N$ matrix describing
the connection between the $\alpha$th lead and the vertices $n=1,2$ with matrix elements $W_{\alpha,n}=\delta_{\alpha,n}$. The $N
\times N$ matrix $M$
\begin{align} \label{Mmatrix}
	M_{nm} =
	\begin{cases}
		\lambda_n k-\sum_{l\neq n}{\cal A}_{nl}\cot kL_{nl}, \quad n = m\\
		{\cal A}_{nm}\csc kL_{nm}, \quad \quad \quad \quad \quad \,\,\, n \neq m
	\end{cases}
\end{align}
incorporates information about the metric and the connectivity of the graph, where ${\cal A}$ is the adjacent matrix having elements
zero (whenever two vertices are not connected) and one (whenever two vertices are connected) \cite{kottos2000chaotic}. The constant
$\lambda_{n}$ characterizes the linear dielectric properties of the vertices and can be in general complex in order to take into account
losses. The wavenumber of the propagating wave is $k=\omega n_r/c$ where $\omega$ is its angular frequency and $n_r$ is the index
of refraction of the coaxial cable, while $c$ is the speed of light. Finally, $(M_{NL})_{n,m} = k f(|\phi_N^{(\alpha)}|^2)\delta_{nm}
\delta_{n,N}$ incorporates the nonlinearity at the $n=N$ vertex. In this work we will be mainly considering Kerr or saturable nonlinearities.

Using Eq.~(\ref{GMM}) we find that the field intensity at the nonlinear vertex ${\text x}_{\alpha}= |\phi_N^{(\alpha)}|^2$ is a root of
the equation (see SM Sec.~\ref{Appendix-GGF})
\begin{align}
\label{cardano}
{\text x}_{\alpha}\left[\left|b\right|^2+\left|kf({\text x}_{\alpha})\right|^2-2{\cal R}\bigl(kb^*f({\text x}_{\alpha})\bigr) \right]=
4\left|A_{\alpha} c_{\alpha}\right|^2\,,
\end{align}
where {\it the coefficients $b$ and $c_{\alpha}$ depend on the properties (metric and connectivity) of the linear graph}. In addition,
{\it $c_{\alpha}$ incorporates the information about the vertices $n=1,2$ which are connected with the leads $\alpha=1,2$}. Further
manipulations allow us to turn Eq.~(\ref{cardano}) to a cubic algebraic equation for ${\text x}_{\alpha}$ which can be solved exactly
using Cardano's formula (see SM Sec.~\ref{Appendix-GGF}). Substituting the value of ${\text x}_{\alpha}$ back in Eq.~(\ref{GMM})
allows us to evaluate the rest of the components of the scattering vector field $\Phi^{(\alpha)}$. Specifically, the field amplitude
$\phi^{(\alpha)}_{n_\beta}$ associated with the vertex $n_{\beta}\neq \alpha$ is
\begin{align} \label{trans}
\phi^{(\alpha)}_{n_\beta} = 2iA_{\alpha}\left[q_{\alpha\beta}-\frac{c_{\alpha}c_{\beta}}{b-kf({\text x}_{\alpha})} \right]\,,
\end{align}
where the constant $q_{1,2}=q_{2,1}=q$ encodes information about the structure (metric and connectivity) of the graph and the vertices
where the TLs are attached (see SM Sec.~\ref{Appendix-GGF}). At the same time, the continuity condition at the
vertex $n$ enforces that the transmitted wave has the same amplitude given by Eq.~(\ref{trans}). Consequently, the transmittance is
$T_{\alpha} \equiv \left|\frac{\phi^{(\alpha)}_{n_\beta}}{A_{\alpha}}\right|^2$. For real-valued $f(|\phi_N^{(\alpha)}|^2)$, the transmittance
takes the simple form
\begin{equation} \label{genT}
T_{\alpha}=4|q|^2\frac{[X_{\alpha} - \Re(\frac{c_1c_2}{qk\Im(\frac{b}{k})})]^2 + [1 - \Im(\frac{c_1c_2}{qk\Im(\frac{b}{k})})]^2}
{X_{\alpha}^2 + 1}\,,
\end{equation}
where $X_{\alpha} = \frac{\Re(\frac{b}{k})-f({\text x}_{\alpha})}{\Im(\frac{b}{k})}$.
(see SM Sec.~\ref{appendix:TFL1} for a generalization to complex-valued nonlinearities).

\begin{figure*}[htp!]
\centering
\includegraphics[width=1\linewidth]{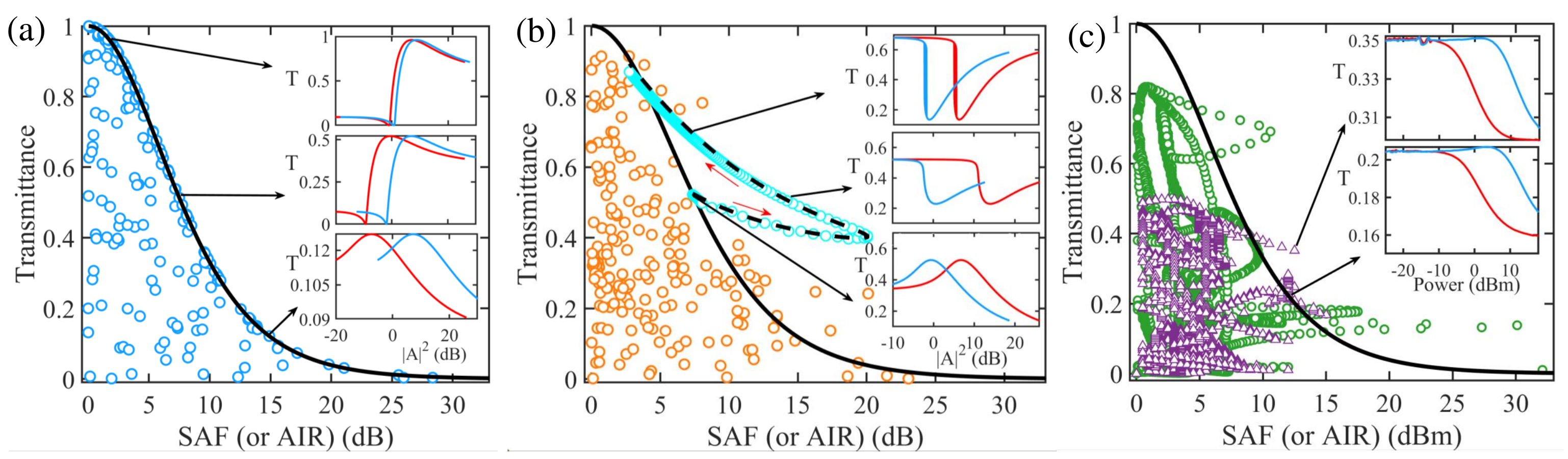}
\caption{{\bf Transmittance and transmittance bounds versus structural asymmetry factor (\SAF{}) or asymmetric intensity range (\AIR{}).}
(a) Lossless graph. The insets show the transmittances versus input intensity from each of the two leads (red and blue lines) for three different
\SAF{} graph configurations. (b) Lossy graph with losses on node $n_0=3$. The light blue circles indicate maximum transmittance for a graph
configuration with increasing loss (along the direction of the red arrow) on node 3. The insets correspond to different losses for a fixed graph configuration.
(c) Measurements (purple
triangles) and simulations (green circles) for a graph with bond-losses and a lossy saturable nonlinearity. The insets show measurements
corresponding to the same \SAF{} but different maximum transmission values. The black solid and dashed lines in (a-c) are theoretical predictions
while the colored circles are simulations occurring at various wavelengths and graph configurations. The data acquisition has been performed for
three different graph configurations and for a frequency range $\nu\in [6.1GHz,6.5GHz]$ with resolution of $\delta \nu=0.4MHz$.
}
\label{fig2}
\end{figure*}

Equations~(\ref{cardano},\ref{genT}) indicate that $T_1\neq T_2$ for two incident waves with the same amplitude $A_1=A_2$ and fixed
wavenumber $k$ that are injected from ports $\alpha=1$ or $2$, whenever the scattering field intensities at the position of the nonlinear vertex
are different from one another, i.e.\ ${\text x}_{1}\neq {\text x}_2$. In this case $X_1\neq X_2$ leading to different transmittances. It is
important to highlight that this non-reciprocal response does not require any form of external bias: the excitation field itself acts as a bias
and triggers the system into a ``high-transmission'' or ``low-transmission'' state depending on incident TL. The experimental results for the
asymmetric transmission due to the presence of a nonlinear vertex is shown in subfigure Fig.~\ref{fig1}d. These measurements are compared
with the results from the graph modeling Eqs.~(\ref{GMM},\ref{Mmatrix}) which are shown in Fig.~\ref{fig1}e. Although the agreement
between theory and experiment is nice, a further refined modeling that takes into consideration the resonant nature of the nonlinear vertex
(see SM Sec.~\ref{Appendix-Hybrid-Model}) provides an even better description of the asymmetric transport, see Fig.~\ref{fig1}f. Whenever
this latter approach is used below, we will refer to it as resonant-graph modeling.

A further analysis of Eq.~(\ref{cardano}) allows us to identify the amplitude range for which asymmetric transport occurs. Specifically,
from the right-hand-side of this equation we conclude that the scattering field intensity ${\text x}_{\alpha}$ at the nonlinear vertex is the
same for a left ($\alpha=1$) and a right ($\alpha=2$) incident waves as long as they satisfy the relation $\left|A_{1} c_{1}\right|^2= \left|
A_{2} c_{2}\right|^2$. The latter equality shows that the field intensity at the nonlinear vertex ${\text x}_{\alpha}$, and therefore the
nonlinear electric potential, from port 2 is equal to the one from port 1, if the input power from port 1 is $\SAF\equiv \left|\frac{c_2} {c_1}
\right|^2$ times larger than that from port 2. Given that the same field intensity ${\rm x}_{\alpha}$ from different ports implies the same
transmission coefficient, we deduce that transmission from different ports is the same if the input power from port 1 is $\SAF$ times larger
than from port 2. The ratio of these input powers that lead to the same transmission defines the $\AIR{}\equiv \max\Bigl\{\left|\frac{A_1}
{A_2}\right|^2;\left|\frac{A_2}{A_1}\right|^2\Bigr\}$ (see Fig.~\ref{fig1}c). Within the \AIR{}, the graph largely breaks Lorentz reciprocity,
since the transmission levels in opposite directions are markedly different for the same input power and frequency. It follows that the \AIR{}
is equal to the \SAF{}, i.e. $\AIR=\SAF$.

\noindent{\bf Bounds for Transmission Asymmetry --} The maximum transmittance can be used as an upper bound for the transmission
asymmetry since $T\geq 0$ in all cases and, therefore, $\dTmax=\Tmax-T_\mathrm{min}\leq \Tmax$.

From Eq.~(\ref{genT}) we derive an upper bound for the transmittance by maximizing $T_{\alpha}$ with respect to $X_{\alpha}$.
For real-valued nonlinearities we have
\begin{align}\label{Eq:Tmax-lossy}
	\Tmax &= 2|q|^2\left(\left|\Lambda\right| \sqrt{\left|\Lambda\right|^2 + 4\left[1 -
	\Im{\left(\Lambda\right)}\right]} \right.\nonumber\\
	&\quad\quad\quad\quad+ \left.\left[\left|\Lambda\right|^2 + 2\left(1 - \Im{\left(\Lambda\right)}\right)\right]\right)\,
\end{align}
where $\Lambda=\frac{c_1c_2}{qk\Im(\frac{b}{k})}$ (for a more general case of complex nonlinearities see SM Sec.~\ref{appendix:TFL1}).
Equation (\ref{Eq:Tmax-lossy}), together with Eq. (\ref{Eq:Tmax-lossy-all-App}) of the supplement, are the main results of this paper. They
provide a guidance on the dependence of AT on the parameter $\Lambda$ which encodes the structural characteristics of the graph.

The special case of  lossless graphs, can be also retrieved from the above expression and occurs when $\Im{\left(\Lambda\right)}=1$ (see
SM Sec. \ref{Appendix-proof}). In this case, Eq. (\ref{Eq:Tmax-lossy}) simplifies to
\begin{align}\label{Eq:T-max}
	\Tmax = \frac{4\cdot \text{\SAF}}{(\text{\SAF} + 1)^2}.
\end{align}
This expression is nicely confirmed from our numerical data for a lossless graph with Kerr (open
blue circles) and saturable nonlinearities (not shown) in Fig.~\ref{fig2}a. A further investigation reveals that there is an interlinked
relation between the maximum transmittance achieved for a specific incident power and the \SAF{} (or equivalently of the \AIR{} \cite{cotrufo2021nonlinearity1,cotrufo2021nonlinearity2,sounas2018fundamental}). This is reflected in the three examples shown in
the inset of Fig.~\ref{fig2}a, where we report the transmittances $T_1, T_2$ associated with the same incident wave being injected
from channels $\alpha=1$ and $\alpha=2$, respectively, versus the incident power. We find that an increase in the \AIR{} (or equivalently
in the \SAF{}) is associated with a decrease of the maximum transmittance and vice-versa as expected by Eq. (\ref{Eq:T-max}).

Equation~(\ref{Eq:T-max}) has been previously derived as the upper bound of nonlinear AT. Its derivation assumed non-linear Fano
resonators with time-reversal symmetry (i.e. no losses) and has utilized the coupled-mode theory (CMT) framework \cite{sounas2017time,
sounas2018fundamental,cotrufo2021nonlinearity1}. Here, however, we have derived Eq.~(\ref{Eq:T-max}) for an actual nonlinear chaotic
system, where {\it $\SAF{}$ explicitly refers to specific bulk asymmetries pertaining to the topology and metrics of the graph}. Taken the
technological importance of AT, it is natural to investigate and establish (necessary) conditions which enforce the violation of Eq. (\ref{Eq:T-max})
and allow for an enhanced \AIR{} (for a fixed \Tmax{}) or enhanced transmission asymmetry bound (for a fixed \AIR{}) given by Eq.
(\ref{Eq:Tmax-lossy}).

As discussed above, Eq.~(\ref{Eq:T-max}) does not hold when losses are introduced in the system. However, the lossy elements need
to be strategically placed either on the bonds of the graph or at vertices not connected to the two TLs or the nonlinear vertex, i.e.
$n_\mathrm{loss}\neq 1,2, N$ (see SM Secs. \ref{Appendix:LossOnLeadVertex},\ref{Appendix-proof1},\ref{Appendix-proof}). In the
opposite case of losses located at the non-linear vertex, a simple renormalization
of the non-linearity (so that it incorporates the absorption term) results to an upper bound given by Eq.~(\ref{Eq:T-max}). Similarly, when
the losses are implemented on a vertex connected to the TLs, a new bound is found which is a stricter version of Eq. (\ref{Eq:T-max}) (see
SM Sec.~\ref{Appendix:LossOnLeadVertex}). The interferences between, at least, two nearby resonance modes can result in a
violation of Eq. (\ref{Eq:T-max}) much alike in case of AT due to the presence of a magnetic field \cite{TRSV} (see SM Sec. \ref{TRSV}).
Finally, from Eq. (\ref{Eq:Tmax-lossy}) we speculate that if $\Im{\left(\Lambda\right)}<1$, the lossy graph configurations might violate
the lossless bound Eq. (\ref{Eq:T-max}). Detail numerical analysis has confirmed that the above inequality is a necessary but
not sufficient condition for violating the lossless limit (see SM Sec. \ref{TRSV}).

A numerical example where the violation of Eq.~(\ref{Eq:T-max}) occurs for a tetrahedron graph with losses at the vertex $n_0=3$, is
shown in Fig.~\ref{fig2}b. Such targeted arrangement of loss, is effectively equivalent to a new graph configuration, where a third
(fictitious) channel is attached to the node $n_0$ thus changing the topology of the graph and affecting indirectly the coupling between
this vertex and the other vertices. While Eq.~(\ref{Eq:T-max}) is violated for intermediate values of loss, it is still respected in the two
limiting cases of zero and very large losses at the $n_0$-vertex. The second limit is understood as an impedance-mismatch phenomenon:
due to the large imaginary ``electric potential'', the $n_0$-vertex is decoupled from the rest of the graph which now acts as a lossless system
with $N-1$ vertices and thus it again satisfies the bound of Eq.~(\ref{Eq:T-max}). In Fig.~\ref{fig2}b we demonstrate the trajectory of
the maximum transmittance versus \AIR{} as the losses at the vertex $n_{0}=3$ of a tetrahedron graph increases. The numerical data
(light blue cycles) for \Tmax{} are nicely matching the theoretical results (dashed black line) of Eq.~(\ref{Eq:Tmax-lossy}) indicating
that the deterioration of \Tmax{} for increasing losses occurs at a slower rate than the enhancement of $\AIR{}$. At some loss-strength,
the \AIR{} reaches its maximum value. Further increase of loss results in a decrease (increase) of \AIR{} (\Tmax) towards its
``impedance-mismatch'' limit.

At Fig.~\ref{fig2}c we report our measurements (purple triangles) for the graph of Fig.~\ref{fig1}, with uniformly distributed losses at
the bonds of the graph. A violation of Eq.~(7) is evident and it is further supported from our simulations (green cycles) using a resonant
-graph modeling. The insets in Fig.~\ref{fig2}c report the experimental transmittances $T_1, T_2$ for two cases with the same \SAF- the
upper one exceeds the bound, while the lower case corresponds to a configuration that respects the bound (see black arrows).

\begin{figure*}
\centering
\includegraphics[width=0.85\linewidth]{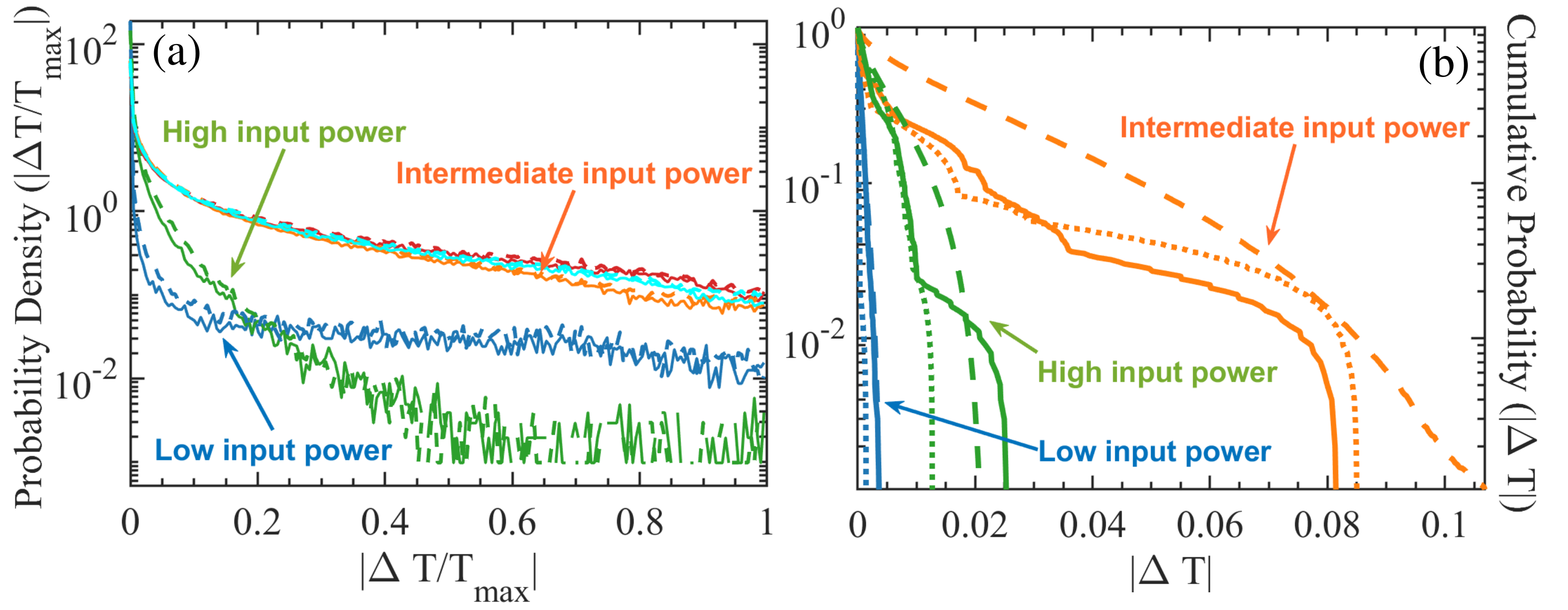}
\caption{\label{fig3}
{\bf Transmission asymmetry distribution for different input powers}. (a) Probability density distribution of transmission asymmetry (normalized
by maximum transmission corresponding to certain \AIR{}) for a tetrahedron graph model with Kerr nonlinearity in one node. The blue, orange
and green solid lines correspond to \AIR{}=4 and input amplitudes 0.1, 10, 1000 [a.u], respectively. The blue, orange, and green dashed lines are
the corresponding RMT results with equivalent input amplitudes 0.0067, 0.44, 33.31 [a.u] (arbitrary units). The light blue and red solid lines are
the results for graphs with \AIR{}=8 and 16, respectively and input amplitude 10 [a.u.], while the corresponding RMT results are shown as dashed
lines of the same color. (b) The cumulative probability distribution of transmission asymmetry for the tetrahedron graph of Fig.~\ref{fig1}a with a
saturable nonlinearity at vertex $N=4$. The blue, orange and green solid lines indicate measurements with input power -25, 1, 17\,dBm respectively.
The dotted lines are the results from the resonant-graph model with the same input power as in the experiment. The dashed lines are the corresponding
RMT modeling with input power -34.4, -7.7, 9.4\,dBm.
}
\end{figure*}

\noindent{\bf Universal Statistics for Transmission Asymmetry--} Motivated by the success of RMT in describing statistical properties of linear wave-
chaotic scattering systems we postulate here an ansatz that the distribution of transmission asymmetries when rescaled with \Tmax{}, i.e.\ ${\cal P}_A
(\Delta T/T_{\rm max})$, for a fixed incident amplitude $A$, is universal. The RMT approach assumes that the chaotic scattering set-up is modeled by
an ensemble of $N\times N$ symmetric matrices $H=H^T$ with random elements taken from a Gaussian distribution with zero mean and standard
deviation $\sigma_{ij}=1/2\pi \sqrt{N}$. As in the case of graphs settings, we assume a monochromatic incident wave with a frequency $\omega$
and amplitude $A_{\alpha}$ injected in one of the two ports $\alpha=1,2$. The ports are coupled to the scattering domain with a coupling strengths
$w_{\alpha}$. The steady-state CMT equations that describe the scattering process are
\begin{align}
&(\omega - H_{eff} - H_{NL})\Phi^{(\alpha)} = iW^TI^{(\alpha)},\label{Eq:MCMT-1} \\
&\boldsymbol{O}^{(\alpha)} = C\boldsymbol{I}^{(\alpha)} + W\Phi^{(\alpha)} \label{Eq:MCMT-2}\,,
\end{align}
where $\Phi^{(\alpha)}$, $I^{(\alpha)}$ and $O^{(\alpha)}$ are the scattering, incident and outgoing vector fields respectively. The effective
Hamiltonian $H_{eff}= H-  i\frac{W^TW}{2}$ describes the wave dynamics in the (linear) complex scattering domain when it is coupled to
ports while $(H_{NL})_{nm}=f(|\phi_N^{(\alpha)}|^2)\delta_{nN}\delta_{nm}$ describes the non-linear interactions affecting the $N-$th resonant
mode. The system-ports coupling is described by the matrix $W$ with elements $W_{n,\alpha}=\delta_{n,\alpha}w_{\alpha}$ ($n=1,\cdots,N$).
By solving for $\Phi^{(\alpha)}$ from  Eq.~(\ref{Eq:MCMT-1}) and substituting into Eq.~(\ref{Eq:MCMT-2}), we get $\boldsymbol{O}^{(\alpha)}
=S\boldsymbol{I}$ where
\begin{align}
S= \big[-\mathbf{1} + iW(\omega - H_{eff} - H_{NL})^{-1}W^T \big]\,,
\label{SCMT}
\end{align}
is the ${\text x}_{\alpha}=|\phi_N^{(\alpha)}|^2$-dependent scattering function $S$. Similar to the case of graphs, ${\text x}_{\alpha}$ is a
solution of an algebraic equation that depends on $A_{\alpha}$, and therefore, $S=S(A_{\alpha})$.

An appropriate RMT modeling requires to supplement our scheme with two additional inputs (for details see SM Sec.~\ref{appendix-RMT}.
The first one involves the values of the coupling elements $w_1,w_2$  such that the RMT modeling takes into account system-specific direct
processes occurring at graphs. The latter are encoded in the energy (or ensemble) averaged $S-$matrix. A direct comparison between the RMT
and the graph scattering matrix in the linear domain gives $w_\alpha = \sqrt{\frac{1}{\pi}\frac{1 - |\langle S_{\alpha,\alpha}\rangle|}{1 +
|\langle S_{\alpha,\alpha}\rangle|}}$ \cite{FS97,KS03}. The second information is the appropriate RMT modeling of the nonlinear coefficients
that define the nonlinearity strength. Equivalently, we identify the incident field amplitudes for which the RMT and the graph model, lead to a
statistically equivalent nonlinear term. By comparing the scattering functions of the graph and the RMT (see Eq.~(\ref{Sgraph}) and
Eq.~(\ref{SCMT}), respectively) we get
\begin{align}
\label{Eq:NLEqCMTGraph}
\frac{2f_{RMT}(\langle|\phi_N^{RMT}|^2\rangle)}{w_1^2} = f_G(\langle|\phi_N^G|^2\rangle)\,,
\end{align}
where $w_1$ is found from above. Expressing $\phi_N^{RMT}, \phi_N^G$ in terms of $A_{\alpha}^{RMT}, A_{\alpha}^{G}$ allows us to
establish an equivalence between the incident fields of the RMT and graphs models that produce the same nonlinear effects.

In  Fig.~\ref{fig3}a, we report the probability density distribution for the rescaled transmission asymmetry ${\cal P}(\left|\widetilde{\Delta T}
\right|=\left|\Delta T\right|/T_{\rm max})$ for different \AIR{}=4, 8, 16 and input amplitudes, $A_{\alpha}$=0.1 (0.007), 10 (0.44), 1000 (33.3)
for the graph (solid lines) and the equivalent RMT system (dashed lines). For the purpose of the analysis, we have used a Kerr-type nonlinearity.
The agreement between them confirms the applicability of RMT modeling to describe the statistical properties of transmission asymmetries.
Furthermore, the various distributions are weakly dependent on the value of \AIR{} for fixed incident powers while they differ dramatically for
different $A$-values
(and fixed \AIR{}).For low incident powers (negligible nonlinear effects) the distribution is concentrated around the origin, signifying that the
asymmetry is essentially suppressed. As the incident power increases the variance of the distribution is acquiring a maximum value reflecting
a large transmission asymmetry. Further increase of the incident power leads to a suppression of the variance and the distribution is again
concentrated near the origin. The revival of the symmetric transport for high incident powers is associated with an impedance-mismatching
phenomenon that leads to an effective decoupling of the nonlinear vertex due to the high values of the nonlinear electrical potential. As a result,
the system acts again as a linear one of $N-1$ vertices, i.e.~reciprocity is restored. The same non-monotonic behavior of the variance of ${\cal P}
(\left|\Delta T\right|)$ occurs also for saturable nonlinearities; albeit the physical mechanism for the reciprocity revival at high powers is different.
Namely, it is associated with the saturable nature of the nonlinearity which above a critical incident power acquires a fixed (saturable) value.

In Fig.~\ref{fig3}b we report integrated transmission asymmetry distribution evaluated from our experimental results (solid lines) for the graph
shown in Fig.~\ref{fig1}a together with the results of the resonant-graph modeling (dotted lines) and the calculations from RMT modeling (dashed
lines) Eqs.~(\ref{Eq:MCMT-1},\ref{Eq:MCMT-2}). In these calculations we have used a saturable nonlinearity that describes our hybrid diode-
resonator system (see SM Secs.\ref{Appendix-Para-Fitting},\ref{Appendix-Hybrid-Model}). An overall nice agreement between measurements,
resonant-graph modeling and RMT re-confirms the validity of our assumption. Specifically, we are able to observe in all cases the same non-
monotonic trend of the $\Delta T$-support of the integrated distribution function as the incident power increases. The smoother behavior of the
integrated transmission asymmetry in case of the RMT modeling is attributed to an additional averaging over different realizations of the
Hamiltonian $H$, which has not be done in the experiment neither in the graph modeling.

\section*{Conclusions}

We have established, experimentally and theoretically, a statistical description of the asymmetric transport (AT) occurring due to the interplay of
nonlinearity with wave-chaos. Our platform consisted of a prototype chaotic system -- a non-linear microwave complex network of coaxial cables
(graphs). The simplicity of this model allowed us to find an {\it exact expression for the upper bound of AT} irrespective of the presence/absence
of losses or resonant coupling conditions. The special case of lossless graphs is treated as a limit of the general expression and reproduces previously
known results \cite{sounas2018fundamental}. Our results connect the AT with the structural asymmetry factor (\SAF) that characterizes the underlying
linear graph. The latter dictates the asymmetric intensity range (\AIR) over which the nonlinear graph demonstrates AT. The simplicity of the model
allowed us to establish (necessary) conditions for enhanced \AIR{} (for a fixed \Tmax{}) or enhanced transmission asymmetry bound (for a fixed
\AIR{}) with respect to previous predictions that were referring to lossless systems. Our conclusions have been confirmed by a nonlinear RMT
modeling which describes the universal statistical features of transmission asymmetries $\Delta T$ of a typical nonlinear chaotic cavity. Using the
RMT-description, we established a non-monotonic behavior of the broadening of the probability distribution of the transmission asymmetries
$\Delta T$ which agrees with our experimental findings with microwave graphs. We find that for weak and strong incident powers the distribution
shrinks around the origin $\Delta T=0$ signifying symmetric transport. Instead, at some intermediate value of the incident power the distribution
acquires its maximum spread. This behavior is a direct consequence of an impedance mismatch phenomenon which decouples the nonlinear element
from the complex surrounding system, similar to the interplay of super-radiance and resonance trapping.

\section*{METHODS}
\noindent{\bf Experimental Implementation and Characterization of the Non-Linear Vertex --}
In our experiment, we implement it by substituting the Tee-junction with a cylindrical resonator (ceramics ZrSnTiO with permittivity $\epsilon\approx
36$, height 5\,mm, diameter 8\,mm, resonance frequency around $\nu_0 \approx 6.885$ GHz and a line width $\gamma \approx 1.7$ MHz) which is
inductively coupled to a metallic ring (diameter 3\,mm) that is short circuited to a diode (detector Schottky diode SMS 7630-079LF, from Skyworks),
see inset of Fig.~\ref{fig1}a. As a result the $z$-directional magnetic field at the resonator of the transmitted signal is inductively coupled to the fast diode.
The strength of the magnetic field dictates the value of the current at the ring and consequently the voltage across the diode. The latter defines the state
of the diode: the “on” state is associated with high voltage (high incident power) and leads to high nonlinearities; the “off” state is associated with low
voltage (low incident power) and leads to low nonlinearities. The nonlinear resonator is designed to operate at 6.1-6.5\,GHz. It is coupled with the rest
of the graph via ``kink'' antennas. The system is coupled to external transmission lines (coaxial cables) attached to $n=1, 2$ nodes of the graph thus
changing their valency to ${\tilde v}_{n=1,2}=v_n+1$. Each transmission line supports a single propagating mode and it is connected to one port of
the Vector Network Analyzer (VNA). This type of resonator nonlinearity has been already used to realize topological limiters in a coupled resonator
framework \cite{jeo20}.

\noindent{\bf Properties of the coaxial cables --}
The lengths of the coaxial cables that have been used for the experimental implementation of the graph of  Fig. \ref{fig1}a: are $L_{12}=680$ mm,
$L_{23}=599$ mm, $L_{13}=277$ mm, $L_{24}=359$ mm, $L_{34}=230$ mm,  $L_{14}=433$ mm. The electrical permitivity of the coaxial
cables has been extracted via best fit of the transmittances/reflectances with the expressions derived from the theoretical analysis of a tetraherdon
structure and was found to be $\epsilon\approx 1.56(\pm0.07)+i0.0015(\pm0.0005)$. For consistency, we have also analyzed the transmission/reflection
from a single cable when connected to a VNA and found similar values of the electrical permittivity of the wires.

\noindent{\bf Mathematical Modeling Using Graph Theory --}
The theoretical analysis, assumes that the length of each bond $l_B=l_{\bar B}$ is taken from a box distribution centered around some mean value
${\bar l}$, i.e., $l_B\in [{\bar l} -W_B/2,{\bar l}+W_B/2]$. The position on bond $B$ is defined as $x_B\equiv x_{nm}$, with $x_{B}=0(l_B)$
on vertex $n(m)$, thus $x_{\bar B}\equiv x_{mn} = l_B - x_{nm}$. The scattering field on the bonds satisfies the Helmholtz equation
\begin{equation}\label{Helm}
	\left(\frac{d^2}{d x_{B}^2}+ k^2 +k^2\left(\lambda_n+\delta_{n N}f(|\phi_{N}^{(\alpha)}|^2)\right)\delta(x_{B})\right)\psi_{B}^{(\alpha)}= 0\,,
\end{equation}
where $\psi_{B}^{(\alpha)}(x_{B})$ is the electric potential difference at position $x_{B}$, $k=\omega n_r/c$ is the wavenumber of the propagating
wave with
frequency $\omega$, $n_r$ is the relative index of refraction of the coaxial cable, $c$ is the speed of light, $\lambda_n$ is the dielectric coefficient
at node $n$, $\delta_{nN}$ is the Kronecker delta function, and the superscript $\alpha = 1, 2$ indicates the lead from which the incident wave has
been injected. In this formulation, the losses in the coaxial cables are modeled by a complex-valued refraction index $n_r$ while losses at the vertices
are modeled by complex $\lambda_n$. The scattering field $\psi_{B}^{(\alpha)}(x_{B})$ can be expressed in terms of its value at the vertices
$\psi_{nm}^{(\alpha)}(x_{nm}=0)=\phi_{n}^{(\alpha)}$ and $\psi_{n,m}^{(\alpha)}(x_{nm}=l_b)=\phi_{m}^{(\alpha)}$. It is, therefore, useful
to introduce the scattering vector field $\Phi^{(\alpha)} = (\phi_1^{(\alpha)},\phi_2^{(\alpha)},\cdots, \phi_{N}^{(\alpha)})^T$. Finally,
$f(|\phi_{N}^{(\alpha)}|^2)$ is the nonlinear dielectric coefficient associated with vertex $N$. For Kerr nonlinearity, we have $f(|\phi_{N}^{(\alpha)}|^2)
= \chi_K|\phi_{N}^{(\alpha)}|^2$, while for saturable nonlinearity we have $f(|\phi_{N}^{(\alpha)}|^2) = z_1/[1+\chi_s|\phi_{N}^{(\alpha)}|^2]$
with $\chi_K$, $\chi_s$ and $z_1$ being complex parameters.

The wavefunction at any bond $B=(n,m)$ that is connected at a vertex $n$, satisfies the continuity relation $\psi_{B}^{(\alpha)}(x_B=0)=\phi_n^{(\alpha)}$.
It also satisfies the current conservation relation $\sum_{B}^{v_n} \frac{d\psi_{B}^{(\alpha)}}{dx_{B}}|_{x_{B}=0} + \sum_{\mu=1,2} \delta_{\mu,\alpha}\frac{d\psi_{\mu}^{(\alpha)}}{dx}|_{x=0} = -k^2 \delta_{n, N}f(|\phi_{N}^{(\alpha)}|^2)\phi_{n}^{(\alpha)}$, where $\psi_{\mu}^{(\alpha)}$ is
the wavefunction on lead $\mu$.

\bibliography{SSS}
\onecolumngrid
\appendix
\newpage

\begin{center} \Large \bf
	Supplemental Material\\
	for manuscript: Universalities of Asymmetric Transport in Nonlinear Wave Chaotic Systems
\end{center}

\begin{center}
Cheng-Zhen Wang$^1$, Rodion Kononchuk$^1$, Ulrich Kuhl$^2$, and Tsampikos Kottos$^1$

$^1$ Wave Transport in Complex Systems Lab, Department of Physics, Wesleyan University, Middletown, CT-06459, USA\\
$^2$ Universit\'{e} C\^{o}te d'Azur, CNRS, Institut de Physique de Nice (INPHYNI), 06108 Nice, France, EU\\
\end{center}

\renewcommand\thefigure{S\Alph{section}.\arabic{figure}}
\setcounter{figure}{0}
\setcounter{page}{1}
\renewcommand{\thepage}{S\arabic{page}}
\setcounter{page}{1}
\setcounter{equation}{0}
\renewcommand\theequation{S\Alph{section}.\arabic{equation}}
\renewcommand{\appendixname}{Suppl.~material}
\renewcommand\thesection{\Alph{section}}
\vspace{2ex}
\hrule
\vspace{2ex}

\twocolumngrid

\section{Characterization of the Nonlinear Resonator}
\label{Appendix-Para-Fitting}

We have characterized the nonlinear resonator (i.e.\ the form of the saturable nonlinearity $f(|\phi_N|^2)$) and its coupling constants with three kink
antennas by comparing the transmission measurements with the corresponding expressions from a coupled mode theory that describes a three port
scattering set-up (see inset in Fig.~\ref{fig1}a). Two of the ports have been connected with the VNA while in each measurement the third port was
coupled to a 50\,Ohm terminator.

The temporal coupled mode theory that describes this system is:
\begin{align}
i\frac{d}{dt}a(t) &= (\tilde{\omega}-i\frac{WW^T}{2})a(t) + iW|S_+\rangle, \\
|S_-\rangle &= C|S_+\rangle + W^T a(t)\,,
\label{oneres}
\end{align}
where the $1\times 3$ coupling matrix $W=(w_1,w_2,w_3)$ describes the coupling of the resonator with the three kink antennas. In our modeling we
have ignored direct processes between the kink antennas. The angular frequency $\tilde{\omega} = \omega_0 + \Omega$ is expressed as a sum of the
intrinsic angular frequency $\omega_0$ of the resonator, and the saturable nonlinear angular frequency $\Omega = 2\pi (z_0 - \frac{z_1}{1+\chi_s |a|^2})$
associated with the coupling of the resonator with the ring antenna that incorporates the nonlinear diode. The nonlinear frequency shift depends on the
magnetic field intensity $|a|^2$ that induces a current to the ring antenna; thus activating the nonlinear diode. The coupling coefficients $w_n, n=1,2,3$
and the parameters $z_0, z_1$ and $\alpha$ will be treated as fitting parameters (see below).

We proceeded by assuming that an incident harmonic field $|S_+\rangle=S_+e^{-i\omega t}$ generates an outgoing field $|S_-\rangle=S_-e^{-i\omega t}$
at the same frequency. This theoretical assumption has been justified by experimentally confirming that the scattering process does not generate higher
harmonic signals (i.e.\ the outgoing energy is mainly scattered at the fundamental frequency). In this respect we also assume that $a(t)=ae^{-i\omega t}$
is the field amplitude at the resonator. Substitution of the temporal form of the fields in Eq.~(\ref{oneres}) leads to the following equations for the field
amplitudes,
\begin{align}
\omega a &= (\tilde{\omega}-i\frac{WW^T}{2})a + iWS_+, \label{Eq:CMT-1}\\
S_- &= -S_+ + W^T a\label{Eq:CMT-2}
\end{align}
From Eq.~(\ref{Eq:CMT-1}) we solve for the nonlinear steady-state field intensity $|a|^2$.
The nonlinear $3\times 3$ scattering matrix $S$ can be evaluated from Eqs.~(\ref{Eq:CMT-1},\ref{Eq:CMT-2}) by substituting back to them the steady
-state value of $|a|^2$. We get
\begin{align} \label{Sres}
S = -1 + iW^T\frac{1}{\omega - (\omega_0 + 2\pi z_0 - \frac{2\pi z_1}{1+\chi_s |a|^2})+i\frac{WW^T}{2}}W
\end{align}
which can be used for extracting the fitting parameters via comparison with our measurements.

\begin{figure*}
	\centering
	\includegraphics[width=0.9\linewidth]{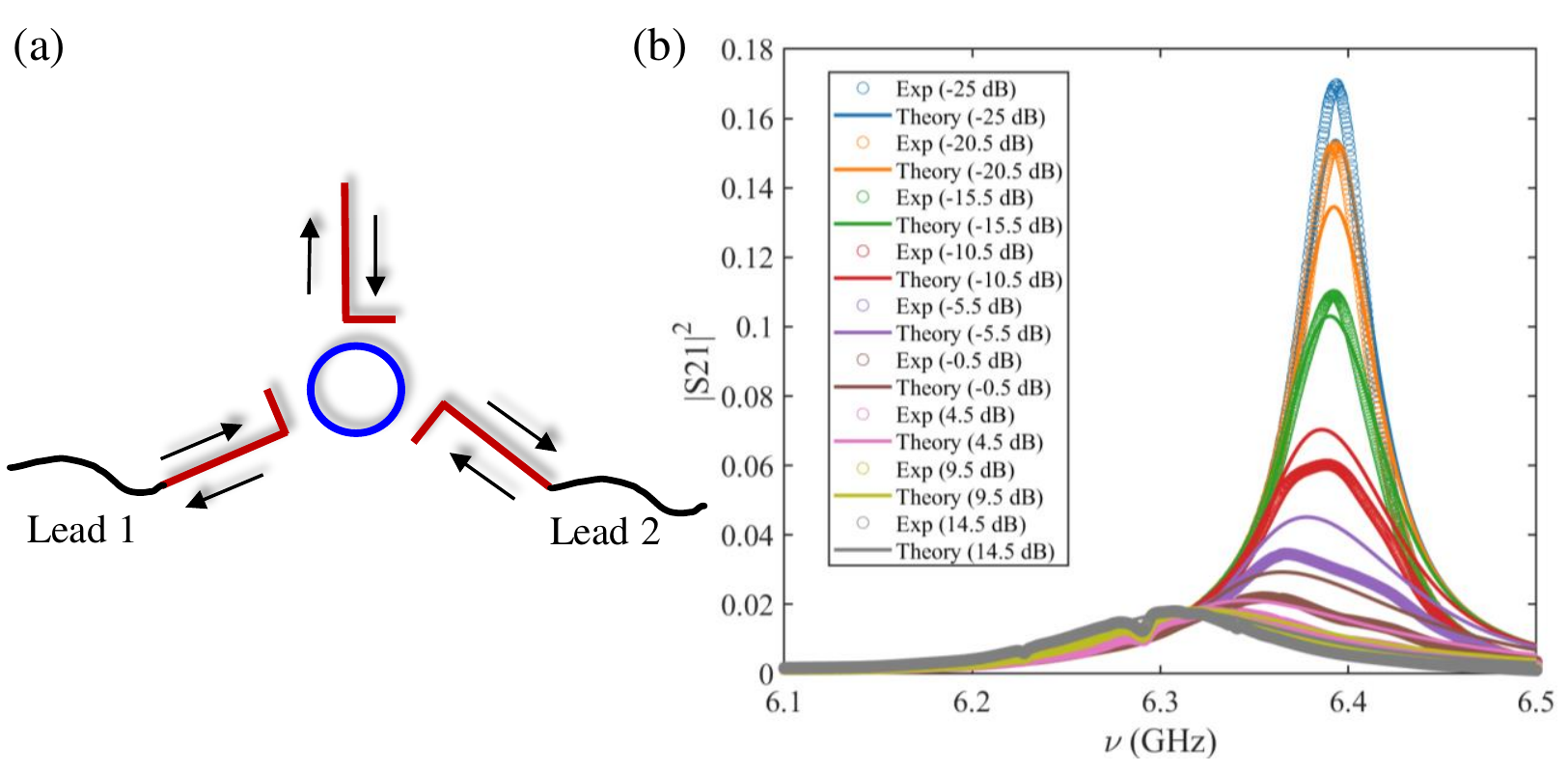}
	\caption{ \label{fig:Fitting-S21} {\bf Experimental implementation and measurements of the nonlinear vertex-}(a) Schematics of the experimental
	set-up used in order to extract the nonlinear fitting parameters. The set-up involves
	three antennas coupled to one resonator. (b) Scattering matrix element $|S_{21}|^2$ versus frequency for different input power. The circles are for
	experimental data, while the solid lines are the best fitting using the theoretical description of Eq.~\ref{Sres}. The extracted best fitting parameters are
	$z_0 = (-86.4 - 59.2i)$ MHz, $z_1 = (-86.4 - 50.0i)$ MHz, $\gamma_1^2=\gamma_2^2=\gamma_3^2 = 62.5$ MHz, $\chi_s = (1.5+1i)\cdot 10^{9}$
	(mW$\cdot$s)$^{-1}$.
		}
\end{figure*}

For weak input powers (e.g.~-25 dBm), $\chi_s |a|^2 \approx 0$. In this case, we can evaluate the transmission from lead $m$ to $n$ as
\begin{align} \label{Eq:fitting-weak-Snm}
  |S_{nm}|^2 = \frac{4w_n^2w_m^2}{[4\pi(\nu - \nu_0 - \Re(z))]^2 + [w_1^2 + w_2^2 + w_3^2 - 4\pi\Im(z)]^2}\,,
\end{align}
where $z = z_0 - z_1$, $\omega = 2\pi \nu$ and $\omega_0 = 2\pi \nu_0$. The maximum value of $|S_{nm}|^2$ is achieved at $\nu_{\rm max} = \nu_0
+ \Re(z)$. Therefore, the experimental evaluation of $|S_{nm}(\nu_{\rm max})|^2$ allows us to extract $\nu_0$ (we consider $\Re(z)= 0$ for simplicity)
together with the $\Im(z) = \frac{(w_1^2 + w_2^2 + w_3^2)}{4\pi} \pm \frac{w_n w_m}{2\pi|S_{nm}(\nu_{\rm max})|}$. Substituting these expressions
into Eq.~(\ref{Eq:fitting-weak-Snm}), allows us to express the scattering cross-section in terms of the coupling coefficients $w_n$ ($n=1,2,3$). The latter
are extracted via a direct fitting with the measured $|S_{nm}(\nu)|^2$ versus $\nu$. This information allows us to evaluate also $\Im(z)$ which needs to
satisfy also the constraint $\Im(z)\leq \frac{w_1^2 + w_2^2 + w_3^2 - w_n^2} {4\pi}$. The latter bound is enforced by the requirement that the reflectance
in the weak incident power limit (which takes the form)
\begin{align} \label{refl}
|S_{nn}|^2 = 1 - \frac{4w_n^2(w_1^2 + w_2^2 + w_3^2 - w_n^2 - 4\pi\Im(z))}{[4\pi(\nu - \nu_0 - \Im(z))]^2 + (w_1^2 + w_2^2 + w_3^2 - 4\pi\Im(z))^2}
\end{align}
must be bounded from above by unity.

Similarly, the analysis of the transmission and the reflection spectrum in the strong input power limit allows us to extract the value of $z_0$. In this case
$\frac{\tilde{z}_1}{1+\chi_s|a|^2} \approx 0$ and Eqs.~(\ref{Eq:fitting-weak-Snm},\ref{refl}) still apply with the modification of $z\rightarrow z_0$. By
repeating the same procedure as previously, we can extract $z_0$. Combining this information with the result for $z = z_0 - z_1$ that we have extracted
from the previous analysis of weak field, we get $z_1$. Finally, the appropriate value of $\chi_s$ has been extracted by using this parameter as a free fitting
parameter for a set of experimental scattering data that we have collected for intermediate values of incident power.

Following the above procedure we find that the best fitting of the experimental data with Eq. (\ref{Sres}) is provided  using the following parameters for
the saturable nonlinerity of the diode $z_0 = (-86.4 - 59.2i)$ MHz, $z_1 = (-86.4 - 50.0i)$ MHz, $\gamma_1^2=\gamma_2^2=\gamma_3^2 = 62.5$ MHz,
$\chi_s = (1.5+1i)\cdot 10^{9}$(mW$\cdot$s)$^{-1}$. Some representative examples of the fitting process are shown in Fig.~\ref{fig:Fitting-S21}.

\section{Graph Formalism}\label{Appendix-GGF}

The wave propagation along a coaxial cable is characterized by the one-dimensional wave equation in the bond connecting vertices $n$ and $m$ given as
\begin{widetext}
\begin{equation} \label{Eq:TEM-equationA}
  \frac{d^2}{d x_{nm}^2}\psi_{nm}^{(\alpha)}(x_{nm}) + \frac{\omega^2\epsilon}{c^2}[\lambda_{n}
  + \delta_{n N} f(|\phi_{N}^{  (\alpha)}|^2)]
   \cdot \delta(x_{nm})\psi_{nm}^{(\alpha)}(x_{nm}) + \frac{\omega^2\epsilon}{c^2}\psi_{nm}^{(\alpha)}(x_{nm}) = 0\,,
\end{equation}
\end{widetext}
where the superscript $\alpha = 1, 2$ indicates the transmission line,  $\epsilon$ is the dielectric constant of the coaxial cables, $\omega = 2\pi \nu$ is the
angular frequency with $\nu$ the microwave frequency and $c$ is the speed of light. The wave number is $k = \sqrt{\epsilon}\omega/c$.
The constant $\lambda_{n}$ characterizes the linear dielectric properties of the vertices (Tee-junctions) and can be in general complex in order to take into
account losses. Finally, $f(|\phi_{N}^{(\alpha)}|^2)$ is the nonlinear dielectric coefficient on vertex $N$ (N=4 in our case). For Kerr nonlinearity, we have
$f(|\phi_{N}^{(\alpha)}|^2) = \chi_k|\phi_{N}^{(\alpha)}|^2$, while for saturable nonlinearity we have $f(|\phi_{N}^{(\alpha)}|^2) = z_1/[1+\chi_s |\phi_{N}^{(\alpha)}|^2]$ with $\chi_k$, $\chi_s$ and $z_1$ being parameters that characterize the nonlinearity. The wavefunction on the vertices is characterized by
the scattering vector field $\Phi^{(\alpha)} = (\phi_1^{(\alpha)},\phi_2^{(\alpha)}, \cdots, \phi_{N}^{(\alpha)})^T$ whose components describe the electric
potential difference at each node $n$.

The wave function on the bonds of the graph are written as
\begin{align} \label{Eq:bond_wave}
\psi_{n m}^{(\alpha)}(x_{n m}) = \phi_{n}^{(\alpha)}\frac{\sin k(L_{n m} - x_{n m})}{\sin kL_{n m}} + \phi_{m}^{(\alpha)} \frac{\sin kx_{n m}}
{\sin kL_{n m}}\,,
\end{align}
while at the leads take the form
\begin{align}\label{Eq:lead_wave}
  \psi_{n}^{(\alpha)}(x) = A_{\alpha}\delta_{n\alpha}e^{-ikx} + \Sigma_{n\alpha}e^{ikx}\,,
\end{align}
where the leads $\alpha = 1, 2$ are connected to the vertices $n = 1, 2$ respectively. The input amplitude from lead $\alpha$ is indicated as $A_{\alpha}$
while $\Sigma_{n\alpha}$ indicates the reflection coefficient (for $n = \alpha$) or transmission (for $n \neq \alpha$) coefficient. We will assume that
$x = 0$ indicates the vertex, where the lead is attached while $x>0$ indicates the outward position in the lead.

At the vertices, the wavefunction must be continuous and must satisfy a current conservation relation.
The wave continuity condition at any bond $b=(n,m)$ that connects vertices $n$ and $m$ reads $\psi_{nm}^{(\alpha)}(x_{nm}=0) = \phi_{n}^{(\alpha)}$.
Similarly, the continuity condition for a wavefunction at the lead $\mu=1,2$ reads $\psi_{\mu n}^{(\alpha)}(x_{n\mu}=0) = \phi_{\mu}^{(\alpha)}$.
The latter relation can be expressed in matrix form as
\begin{align} \label{Eq:Lead_M_cont}
  I^{(\alpha)} + \Sigma^{(\alpha)} = W\Phi^{(\alpha)}\,,
\end{align}
where the $2\times N$ coupling matrix has elements $W_{\alpha,n}=\delta_{\alpha,n}$ while the two-dimensional incident field vector $I^{(\alpha)}$ has
elements $I_{\mu}^{(\alpha)}=A_{\alpha}\delta_{\mu\alpha}$.

The second boundary condition enforces a current conservation at the vertices, and takes the form
\begin{align} \label{Eq:derivative_cont} \nonumber
  \sum_{m} \frac{d\psi_{n m}^{(\alpha)}}{dx_{n m}}(x_{n m}=0) + \sum_{n=1,2}\delta_{n, \alpha}\frac{d\psi_{n}^{(\alpha)}}{dx}(x=0) \\
  = -k^2[\lambda_{n} + \delta_{n, N}f(|\phi_{N}^{(\alpha)}|^2)]\phi_{n}^{(\alpha)}.
\end{align}
Substituting Eq.~(\ref{Eq:lead_wave}) and (\ref{Eq:bond_wave}) into the above Eq.~(\ref{Eq:derivative_cont}), and combining the outcome with Eq.~(\ref{Eq:Lead_M_cont}), we arrive to the following matrix equation for the vector $\Phi^{(\alpha)}$
\begin{align}\label{Eq:Graph-matrix}
  (M + M_{NL} + iW^TW)\Phi^{(\alpha)} = 2iW^TI^{(\alpha)}\,,
\end{align}
where we have
\begin{align}
  M_{nm} =
  \begin{cases}
    -\sum_{l \neq n}A_{n l}\cot kL_{n l} + \lambda_{n}k, \quad n = m\\
    A_{n m}\csc kL_{n m}, \quad n \neq m
  \end{cases}
\end{align}
and $\left(M_{NL}\right)_{NN} = k f(|\phi_N^{(\alpha)}|^2)$ with all other elements to be 0. From Eqs.~(\ref{Eq:Lead_M_cont},\ref{Eq:Graph-matrix}),
we have
\begin{align} \label{Sgraph}
  \Sigma^{(\alpha)} = (-1 + 2iW[M + M_{NL} + iW^TW]^{-1}W^T)I^{(\alpha)} = SI^{(\alpha)}\,,
\end{align}
where $S$ is the scattering matrix which is intensity-dependent in case of nonlinear elements at vertex $N$. Specifically, the matrix $M_{NL}$ depends
on the steady-state value of the scattering field $\Phi^{(\alpha)}$ component at the position of the nonlinear vertex.

The evaluation of the scattering field vector $\Phi^{(\alpha)}$ is done by inverting the matrix $(M+M_{NL}+iW^TW)$ that appears on the left side
of Eq.~(\ref{Eq:Graph-matrix}). Since $M_{NM}$ depends on the wave component $\phi_N^{(\alpha)}$ we are required first to evaluate this field
component. To this end, we first define the $(N-1)\times N$ matrix $G_L$ and the $1\times N$ matrix $G_{NL}$
\begin{align} \label{Eq:G_L-G_NL}
  G_L =
  \begin{pmatrix}
    1 & 0 & \cdots &0 & 0 \\
    0 & 1 & \cdots &0 & 0\\
    \vdots & \vdots & \ddots & \vdots & \vdots \\
    0 & 0 & \cdots & 1 & 0
  \end{pmatrix}, \quad
  G_{NL} =
  \begin{pmatrix}
    0 & 0 & \cdots & 0 & 1
  \end{pmatrix}.
\end{align}
which allow us to separate Eq.~(\ref{Eq:Graph-matrix}) in two sets of equations, namely
\begin{align}
  G_L[M + M_{NL} + iW^TW]\Phi^{(\alpha)} &= G_L 2iW^TI^{(\alpha)}, \label{Eq:Graph-L}\\
  G_{NL}[M + M_{NL} + iW^TW]\Phi^{(\alpha)} &= G_{NL} 2iW^TI^{(\alpha)}. \label{Eq:Graph-NL}
\end{align}
From Eq.~(\ref{Eq:Graph-L}) we get
\begin{align} \label{Eq:Graph-L1}
  G_L\Phi^{(\alpha)} = -\phi_N^{(\alpha)} \big[ H_{N-1} \big]^{-1} {\bf v_N} + 2i\big[ H_{N-1} \big]^{-1} G_LW^TI^{(\alpha)}\,,
\end{align}
where $G_L\Phi^{(\alpha)}=\left(\phi_1^{(\alpha)},\phi_2^{(\alpha)},\cdots,\phi_{N-1}^{(\alpha)}\right)^T$ is a vector that involves the first
$N-1$ components of the vector field $\Phi^{(\alpha)}$, i.e.\ it excludes the field component associated with the nonlinear vertex. The $(N-1)$
-dimensional vector ${\bf v_N}\equiv\left(M_{1N},M_{2N},\cdots,M_{N-1,N}\right)^T$ and the $(N-1)\times (N-1)$ matrix $H_{N-1}=
M_{N-1} + iW_0$ contain complementary information associated with the connectivity of the nonlinear vertex to the rest of the graph and the
characteristic of the ``linear'' part of the network respectfully. The $(N-1)\times(N-1)$ matrix $M_{N-1}$ is
\begin{align}
  M_{N-1} =
  \begin{pmatrix}
    M_{11} & M_{12} & \cdots & M_{1,N-1} \\
    M_{21} & M_{22} & \cdots & M_{2,N-1} \\
    \vdots & \vdots & \ddots & \vdots \\
    M_{N-1,1} & M_{N-1, 2} & \cdots & M_{N-1, N-1}
  \end{pmatrix}
\end{align}
and $W_0=G_LW^TW$ is an $(N-1)\times(N-1)$ matrix with elements $(W_0)_{nm}=\delta_{\alpha n}\delta_{nm}$. From Eq.~(\ref{Eq:Graph-NL}),
we have
\begin{widetext}
\begin{equation} \label{Eq:Graph-NL1}
  kf(|\phi_N^{(\alpha)}|^2)\phi_N^{(\alpha)}
  +
    \begin{pmatrix}
      M_{N1}& M_{N2}& \cdots& M_{N, N-1}
    \end{pmatrix}
    \begin{pmatrix}
      \phi_1^{(\alpha)}\\
      \phi_2^{(\alpha)} \\
      \vdots\\
      \phi_{N-1}^{(\alpha)}
    \end{pmatrix}
    +M_{NN}\phi_N^{(\alpha)} = 0
\end{equation}
\end{widetext}
which, after substituting $G_L\Phi^{(\alpha)}$ from Eq.~(\ref{Eq:Graph-L1}), leads to the following nonlinear equation for $\phi_N^{(\alpha)}$
\begin{align} \label{Eq:Graph-complex}
  -kf(|\phi^{(\alpha)}_N|^2)\phi^{(\alpha)}_N + b\phi^{(\alpha)}_N - 2iA_{\alpha}c_{\alpha} = 0\,.
\end{align}
For a Kerr nonlinearity, $f(|\phi^{(\alpha)}_N|^2) = \chi_k|\phi^{(\alpha)}_N|^2$ the above equation can be written in a cubic form for the intensity
${\rm x}_\alpha = |\phi_N^{(\alpha)}|^2$ at the nonlinear vertex:
\begin{align} \label{Eq:Cubic}
  \left|\chi_k k\right|^2{\rm x}_\alpha^3 - 2\Re(\chi_kkb){\rm x}_\alpha^2 + |b|^2{\rm x}_\alpha - 4A^2_{\alpha}|c_{\alpha}|^2=0
\end{align}
while for a saturable nonlinearity, i.e. $f(|\phi_{N}^{(\alpha)}|^2) = z_1/[1+\chi_s|\phi_{N}^{(\alpha)}|^2]$, we get
\begin{widetext}
\begin{equation}
	|b\chi_s|^2{\rm x}_\alpha^3 + [2\Re(b\chi_s)\Re(b-z_1\chi_s)
	+ 2\Im(b\chi_s)\Im(b-z_1\chi_s)
	- 4|\chi_s|^2|c_{\alpha}|^2|A_{\alpha}|^2]{\rm x}_\alpha^2
	+ [|b-z_1\chi_s|^2 - 8\chi_s|c_{\alpha}|^2|A_{\alpha}|^2]{\rm x}_\alpha
	- 4|c_{\alpha}|^2|A_{\alpha}|^2=0.
\label{saturable}
\end{equation}
\end{widetext}

It is convenient for the further analysis to define the quantities below
\begin{align}
  b &={\bf v_N}^T \big[ H_{N-1} \big]^{-1} {\bf v_N}-M_{NN} \label{Definition-b}\,,\\
  c_{\alpha} &= {\bf v_N}^T \big[ H_{N-1} \big]^{-1} {\bf e_{\alpha}} \label{Definition-c1}\,,\\
  q_{\alpha\beta} &= {\bf e_\alpha}^T \big[ H_{N-1} \big]^{-1}{\bf e_\beta}\,,
\end{align}
where the $(N-1)$-dimensional vector ${\bf e_{\alpha}}$ indicates the coupling with the $\alpha=1,2$ lead and has elements $({\bf e_{\alpha}})_n
=\delta_{\alpha,n}$.

The cubic equations Eq.~(\ref{Eq:Cubic},\ref{saturable}) can be solved using Cardano's formula, that provides the roots of a cubic algebraic equation
of the form
\begin{align} \label{Eq:Phi_N_O}
  a_0 {\rm x}_\alpha^3 + b_0{\rm x}_\alpha^2 + c_0{\rm x}_\alpha +d_0 = 0
\end{align}
with solutions
\begin{align}
  {\rm x}_\alpha^{(1)} &= S + T - \frac{b_0}{3a_0}\,, \label{Eq:Cardano-1}\\
  {\rm x}_\alpha^{(2)} &= -\frac{S + T}{2} - \frac{b_0}{3a_0} + \frac{i\sqrt{3}}{2}(S - T)\,,\label{Eq:Cardano-2}\\
  {\rm x}_\alpha^{(3)} &= -\frac{S + T}{2} - \frac{b_0}{3a_0} - \frac{i\sqrt{3}}{2}(S - T)\,, \label{Eq:Cardano-3}
\end{align}
where $S = \sqrt[3]{R + \sqrt{Q^3 + R^2}}$ and $T = \sqrt[3]{R - \sqrt{Q^3 + R^2}}$, $Q = \frac{3a_0c_0 - b_0^2}{9a_0^2}$ and $R =
\frac{9a_0b_0c_0 - 27a_0^2d_0 - 2b_0^3}{54a_0^3}$. The expression $D = Q^3 + R^2$ is the discriminant of the equation:
If $D>0$, then one root is real and the other two are complex conjugates;
if $D=0$, all three roots are real, and at least two are equal;
if $D<0$, then all three roots are real and unequal.
In the latter case, the system admits bistable solutions.
Based on Cardano's formula, we obtain $|\phi_N^{(\alpha)}|^2$ and from there $\phi_N^{(\alpha)}$.
The other wave components $(\phi_1^{(\alpha)},\phi_2^{(\alpha)}, \cdots, \phi_{N-1}^{(\alpha)})^T$ can be obtained by substituting $\phi_N^{(\alpha)}$ into Eq.~(\ref{Eq:Graph-L1}).

\section{Transmission Formula for Lossless and Lossy Graphs}
\label{appendix:TFL1}

Based on the results of section~\ref{Appendix-GGF} we can determine the transmission in the lossless and lossy graphs. Starting from
Eq.~(\ref{Eq:Phi_N_O}), we can evaluate the field intensity ${\rm x}_{\alpha}= |\phi^{(\alpha)}_N|^2$ at the nonlinear vertex and from
there using Eq.~(\ref{Eq:Graph-complex}) extract the field amplitude as
\begin{align} \label{Eq:phi_N_alpha}
  \phi^{(\alpha)}_N = \frac{2iA_{\alpha}c_{\alpha}}{-kf(\text{x}_{\alpha}) + b}.
\end{align}
Substituting $\phi_N^{(\alpha)}$ into Eq.~(\ref{Eq:Graph-L1}), allows us to evaluate the scattering vector field at all vertices.
Of particular interest is the values of the scattering field at vertex $\mu=1, 2$, where the leads are attached.
In case of incidence waves from the opposite leads $\alpha = 2, 1$, they take the values
\begin{align}
  \phi^{(\alpha)}_{\mu} = p_{\alpha} \phi^{(\alpha)}_N + 2iA_{\alpha} q_{\alpha} = 2iA_{\alpha}\big[\frac{p_{\alpha}c_{\alpha}}{-kf(\text{x}_{\alpha}) + b} + q_{\alpha}\big]\,,
\end{align}
where we have used $p_1 = -c_2$, $p_2 = -c_1$ and $q_{12} = q_{21}=q$.
Subsequently, we can evaluate the transmission as
\begin{align} \label{Eq:T-alpha}
  T_{\alpha} = \frac{|\phi^{(\alpha)}_{\mu}|^2}{A_{\alpha}^2} = 4\left|q-\frac{c_1c_2}{b-kf(\text{x}_{\alpha})}\right|^2\,.
\end{align}
We can further simplify the transmission formula (\ref{Eq:T-alpha}) for the cases, where $k,\chi,z_1$ are complex-values.
These scenarios describe cable losses (due to the complex refractive index) or nonlinear losses (where, however, $\chi_s$ is real). To this end, we introduce a new variable $a=-k\chi$ for Kerr nonlinearities or $a=-kz_1$ for saturable nonlinearities.
This allows us to factorize the nonlinear permittivity as $a\tilde{f}(\text{x}_{\alpha})=-kf(\text{x}_{\alpha})$ with $\tilde{f}(\text{x}_{\alpha})$ being real.
We have
\begin{align} \label{Eq:T-alpha-all-App}
  T_{\alpha} &= 4|q|^2\Big|\frac{a\tilde{f}(\text{x}_{\alpha})+b - \frac{c_1c_2}{q}}{a\tilde{f}(\text{x}_{\alpha}) + b}\Big|^2 \nonumber \\
  &= 4|q|^2\Big|\frac{\tilde{f}(\text{x}_{\alpha})+\frac{b}{a} - \frac{c_1c_2}{qa}}{\tilde{f}(\text{x}_{\alpha}) + \frac{b}{a}}\Big|^2 \nonumber \\
  &= 4|q|^2\Big|\frac{\tilde{f}(\text{x}_{\alpha})+\Re(\frac{b}{a})+i\Im(\frac{b}{a}) - \Re(\frac{c_1c_2}{qa}) - i\Im(\frac{c_1c_2}{qa})}{\tilde{f}(\text{x}_{\alpha}) + \Re(\frac{b}{a}) + i\Im(\frac{b}{a})}\Big|^2 \nonumber \\
  &= 4|q|^2\frac{[\frac{\tilde{f}(\text{x}_{\alpha}) + \Re(\frac{b}{a})}{\Im(\frac{b}{a})} - \Re(\frac{c_1c_2}{qa\Im(\frac{b}{a})})]^2 + [1 - \Im(\frac{c_1c_2}{qa\Im(\frac{b}{a})})]^2}{[\frac{\tilde{f}(\text{x}_{\alpha}) + \Re(\frac{b}{a})}{\Im(\frac{b}{a})}]^2 + 1} \nonumber \\
  &= 4|q|^2\frac{[X_{\alpha} - \Re(\frac{c_1c_2}{qa\Im(\frac{b}{a})})]^2 + [1 - \Im(\frac{c_1c_2}{qa\Im(\frac{b}{a})})]^2}{X_{\alpha}^2 + 1}\,,
\end{align}
where $X_{\alpha} = \frac{\tilde{f}(\text{x}_{\alpha}) + \Re(\frac{b}{a})}{\Im(\frac{b}{a})}$.

For lossless graphs we have that $a\Im(\frac{b}{a})=\Im(b)=b^i$ which allows us to further simplify the above expression for the transmittance.
We have
\begin{align} \label{Eq:Lossless-Tmax-App}
  T_{\alpha}= 4|q|^2\frac{[X_{\alpha} - \Re(\frac{c_1c_2}{qb^i})]^2}{X_{\alpha}^2 + 1}\,,
\end{align}
where we used the equality
\begin{align}\label{Eq:equality-1-App}
\Im(\frac{c_1c_2}{q}) = \Im(b)
\end{align}
for lossless graphs (see proof in section~\ref{Appendix-proof}).

Minimization of the expression Eq.~(\ref{Eq:Lossless-Tmax-App}) with respect to the variable $X_{\alpha}$, give us the minimum transmission
$T_{\alpha}=0$ occurring at $X_{\alpha} = \Re(\frac{c_1c_2}{qb^i})$. Similarly, the maximum value of transmission is
\begin{align} \label{Eq:T-max-App}
  T_{\rm max} &= 4|q|^2\Big[1 + \Re(\frac{c_1c_2}{qb^i})^2\Big]= \frac{4|c_1|^2|c_2|^2}{b^{i2}} \nonumber \\
  &= \frac{4|c_1|^2|c_2|^2}{(|c_1|^2 + |c_2|^2)^2}
\end{align}
and occurs for $X_{\alpha}^{\rm max} = -\frac{1}{\Re(\frac{c_1c_2}{qb^i})}$.
In deriving the latter expression for the maximum transmission we have used another identity for lossless graph (for a proof see section~\ref{Appendix-proof1})
\begin{align} \label{Eq:equality-2-App}
  [\Im(b)]^2 = (|c_1|^2 + |c_2|^2)^2.
\end{align}
Finally, the corresponding field intensity ${\text x}_{\alpha}^{max}$ for which  \Tmax{} occurs, is evaluated by equating the relation for
$X_{\alpha}^{\rm max} = -\left(\Re(\frac{c_1c_2}{q\Im(b)})\right)^{-1}$ with the expression for $X_{\alpha}({\rm x}_{\alpha})$ (see
formula below Eq.~(\ref{genT})).

Using the definition for \SAF{}, being $\SAF = \AIR{} = max\Bigl\{\left|\frac{c_2}{c_1}\right|^2=\left|\frac{A_1}{A_2}\right|^2;\left|\frac{c_1}{c_2}\right|^2=\left|\frac{A_2}{A_1}\right|^2\Bigr\}$, the maximum transmission can be re-written as
\begin{align}
  T_{\rm max} = \frac{4\cdot \text{\SAF{}}}{(\text{\SAF{}} + 1)^2}\,.
\end{align}

Let us finally mention that for a generic graph with losses, the maximum transmission is along the same lines by considering the value of
$X_{\alpha}$ for which $dT_{\alpha}/dX_{\alpha}=0$. Substitution of this value back into Eq.~(\ref{Eq:T-alpha-all-App}), gives
\begin{align} \nonumber
  T_{\rm max} &= 2|q|^2\sqrt{|\frac{c_1c_2}{qa\Im(\frac{b}{a})}|^4 + 4|\frac{c_1c_2}{qa\Im(\frac{b}{a})}|^2[1 - \Im{(\frac{c_1c_2}{qa\Im(\frac{b}{a})})}]} \\
  & + 2|q|^2[|\frac{c_1c_2}{qa\Im(\frac{b}{a})}|^2 + 2(1 - \Im{(\frac{c_1c_2}{qa\Im(\frac{b}{a})})})]
  \label{Eq:Tmax-lossy-all-App}
\end{align}
where the maximum value of transmission is obtained by taking into Eq.~(\ref{Eq:T-alpha-all-App}) the corresponding $X_{\alpha}$ value as
$X_{\alpha}^{\rm max}=\frac{(h_1^2+h_2^2 - 1)-\sqrt{(h_1^2+h_2^2 - 1)^2 + 4h_1^2}}{2h_1}$ with $h_1 = \Re(\frac{c_1c_2}{qa\Im(\frac{b}{a})})$
and $h_2 = [1 - \Im(\frac{c_1c_2}{qa\Im(\frac{b}{a})})]$.

\section{Resonant-Graph Modeling}
\label{Appendix-Hybrid-Model}

An improved modeling of the graph system of Fig.~\ref{fig1} requires to take into account separately the resonant nature of the resonator.
To this end, we have developed a scheme that combines the coupled mode equations~(\ref{Eq:CMT-1},\ref{Eq:CMT-2}) together with the equations that describe the wave propagation in the rest of the graph.

First, we have developed a continuity equation for the wave at the coupling points (kink antennas) between the graph bonds and the resonator based on Eqs.~(\ref{Eq:CMT-2}).
We have
\begin{align}
S_1^+ + S_1^- = \gamma_1 a = \phi_N^{(1)}, \label{Eq:phiN1}\\
S_2^+ + S_2^- = \gamma_2 a = \phi_N^{(2)}, \label{Eq:phiN2}\\
S_3^+ + S_3^- = \gamma_3 a = \phi_N^{(3)}, \label{Eq:phiN3}
\end{align}
where $a$ is the field amplitude at the resonator, $\phi_N^{(m)}$ ($m=1,2,3$) is the wave at the termination point of the coaxial cable (kink antenna), and $S_1^-=I_{1N}e^{-ikL_{1N}}$, $S_1^+=R_{1N}e^{ikL_{1N}}$, $S_2^- = I_{2N}e^{-ikL_{2N}}$, $S_2^+=R_{2N}e^{ikL_{2N}}$, $S_3^- = I_{3N}e^{-ikL_{3N}}$, $S_3^+ = R_{3N}e^{ikL_{3N}}$, with $I_{mN}$ and $R_{mN}$ are the incident and reflected wave coefficients of a wave interacting with the vertex $N$ (resonator) while it is injected from vertex $m$.

As in our previous analysis, we write the wavefunction at each of the bonds (coaxial cables) of the graph as
\begin{align} \label{Eq:Bond-Wave2}
  \psi_{n m}(x_{n m}) = \phi_{n}\frac{\sin k(L_{n m} - x_{n m})}{\sin kL_{n m}} + \phi_{m}\frac{\sin kx_{n m}}{\sin kL_{n m}}.
\end{align}
Similarly, the wave at the leads that connect the graph to the VNA takes the form:
\begin{align}
  \psi_1 = I_1e^{-ikx} + R_1 e^{ikx}\,, \\
  \psi_2 = I_2e^{-ikx} + R_2 e^{ikx}\,.
\end{align}
From the wave continuity relation associated with a vertex that is connected to a lead, we have
\begin{align}
  I_1 + R_1 = \phi_1, \label{Eq:lead1}\\
  I_2 + R_2 = \phi_2.\label{Eq:lead2}
\end{align}
The current conservation condition (refer Eq.~(\ref{Eq:derivative_cont}) in the pure graph derivation) at each of the $N-1$ vertices of the graph (excluding the vertex associated with the nonlinear resonator) can be combined in the following matrix form
\begin{align} \label{Mat}
  (M+iW^TW)\Phi = 2iW^T
  \begin{pmatrix}
    I_1 \\
    I_2
  \end{pmatrix} - aD\,,
\end{align}
where
\begin{widetext}
\begin{equation*}
	M =
	\begin{pmatrix}
	-\sum_{m}A_{1m}\cot kL_{1m} & A_{12}\csc kL_{12} & \cdots & A_{1,N-1}\csc kL_{1,N-1}\\
	A_{21}\csc kL_{21} & -\sum_{m}A_{2m}\cot kL_{2m} & \cdots & A_{2,N-1}\csc kL_{2,N-1} \\
	\vdots & \vdots & \ddots & \vdots \\
	A_{N-1,1}\csc kL_{N-1,1} & A_{N-1,2}\csc kL_{N-1,2} & \cdots & -\sum_{m}A_{N-1,m}\cot kL_{N-1,m}
	\end{pmatrix}
\end{equation*}
and $\Phi = (\phi_1, \phi_2, \cdots, \phi_{N-1})^T$ is the scattering vector field whose components define the value of the field amplitude on each of the $N-1$ vertices (excluding the nonlinear vertex).
Finally, we have defined the vector $D=(\gamma_1\csc kL_{1N}, \gamma_2\csc kL_{2N}, \gamma_3\csc kL_{3N}, 0,\cdots,0)^T$.

From Eq.~(\ref{Mat}) we get
\begin{align}
  \Phi=\begin{pmatrix}
  \phi_1 \\
  \phi_2 \\
  \vdots \\
  \phi_{N-1}
  \end{pmatrix}
  = 2i(M + iW^TW)^{-1}W^T
  \begin{pmatrix}
  I_1 \\
  I_2
  \end{pmatrix}
  -a (M + iW^TW)^{-1}D
\end{align}
\end{widetext}
which allow us to express $\phi_1$, $\phi_2$, $\phi_3$ as a function of the incident wave amplitudes $I_1$, $I_2$ and $a$.
Furthermore, a use of the wave continuity equation Eq.~(\ref{Eq:lead1}, \ref{Eq:lead2}) at the leads allows us to evaluate the reflection amplitutes  $R_1$, $R_2$ in terms of $I_1$, $I_2$, and $a$.
The nonlinear field $a$ is eventually evaluated in terms of input $I_1$, $I_2$ using Eq.~(\ref{Eq:CMT-1}).
Knowledge of the steady-state value of $a$ allows us to evaluate the field dependent scattering matrix and from there the transmittance and reflectance.

To be specific, we can get the wave function on each vertex $n$ (excluding the vertex associated with the nonlinear resonator or vertex associated with the three kink antennas) as
\begin{align}
  \phi_n = \boldsymbol{e}_n\Phi = (a_{n1}, a_{n2})
  \begin{pmatrix}
    I_1 \\
    I_2
   \end{pmatrix}+ a_n a\,,
\end{align}
where $(\boldsymbol{e}_n)$ is an $(N-1)$-dimensional row vector with elements $(\boldsymbol{e}_n)_m = \delta_{n,m}$, $(a_{n1}, a_{n2}) = \boldsymbol{e}_n \cdot 2i(M + iW^TW)^{-1}W^T$ and $a_n = -\boldsymbol{e}_n \cdot (M + iW^TW)^{-1}D$.
Since the scattering field amplitude $a$ at the resonator is unknown, we first solve for $a$.
At the same time one can express the waves on the bonds connected to the resonator using two different forms.
One is given by Eq.~(\ref{Eq:Bond-Wave2}), i.e.
\begin{align} \label{Eq:psi_nN1}
  \psi_{n N}(x_{n N}) = \phi_{n}\frac{\sin k(L_{n N} - x_{n N})}{\sin kL_{n N}} + \phi_{N}^{(n)}\frac{\sin kx_{n N}}{\sin kL_{n N}}\,,
\end{align}
while the other one is
\begin{align} \label{Eq:psi_nN2}
  \psi_{nN}(x_{nN}) = I_{nN}e^{-ikx_{nN}} + R_{nN}e^{-ikx_{nN}}.
\end{align}
Substituting Eq.~(\ref{Eq:psi_nN1}) into Eq.~(\ref{Eq:psi_nN2}), we can get
\begin{align}
R_{nN} = \frac{\gamma_n}{e^{ikL_{nN}} - e^{-ikL_{nN}}}a - \frac{e^{-ikL_{nN}}}{e^{ikL_{nN}} - e^{-ikL_{nN}}}\phi_n\,, \label{Eq:RnN}\\
I_{nN} = -\frac{\gamma_n}{e^{ikL_{nN}} - e^{-ikL_{nN}}}a + \frac{e^{ikL_{nN}}}{e^{ikL_{nN}} - e^{-ikL_{nN}}}\phi_n\,, \label{Eq:InN}
\end{align}
where we have used the relations between $\phi_N^{(n)}$ and $a$ from Eqs.~(\ref{Eq:phiN1},\ref{Eq:phiN2},\ref{Eq:phiN3}).
Finally, by utilizing the formulas (\ref{Eq:RnN}) and (\ref{Eq:InN}) of $R_{nN}$ and $I_{nN}$, we get the expressions for $S_n^+$, $S_n^-$ as a function of $a$, $I_1$, $I_2$.
Substituting $S_n^+$, $S_n^-$ back into Eq.~(\ref{Eq:CMT-1}), we are now able to solve for the field
intensity $|a|^2$ and find the field amplitude $a$ as a function of the input wave amplitudes $I_1$ and $I_2$. By substituting $\phi_1$ and $\phi_2$ (as a function of $I_1$ and $I_2$) into Eq.~(\ref{Eq:lead1}, \ref{Eq:lead2}), we can get the scattering matrix from the relation between the incident and the reflected fields.

\section{Nonlinear Random Matrix Theory Modeling}
\label{appendix-RMT}

\begin{figure*}
\centering
\includegraphics[width=0.85\linewidth]{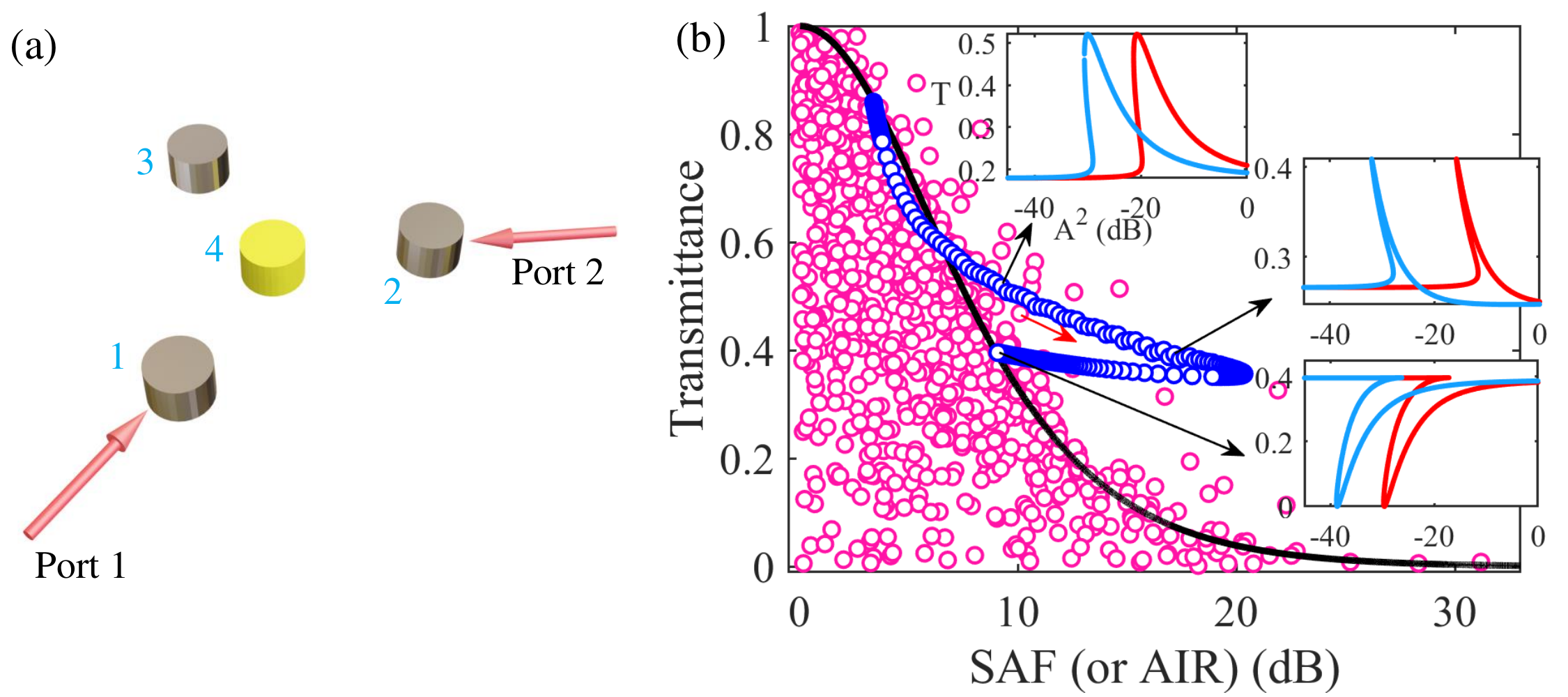}
\caption{{\bf Transmittance bounds using an RMT modeling --} (a) Schematics of the RMT model. (b)Transmittance versus structural asymmetric factor
(\SAF{}) or Asymmetric Intensity Range (\AIR{}) for the RMT model of subfigure \ref{fig5}a. The loss $\Im(\lambda_3)=0.01$ is imposed on mode 3
and the transmittances for an RMT ensemble are marked as pink circles. The blue circles are for one chosen CMT configuration with $\Im(\lambda_3)$
increasing from	0 to 10. The inset shows are transmission versus input intensity corresponding to three different loss values.}
	\label{fig5}
\end{figure*}

For a general system of $N$ modes which are coupled to each other (schematics shown in Fig.~\ref{fig5} (a)), the temporal coupled mode theory (TCMT) that describes the scattering process takes the following form:
\begin{align}
  i\frac{d\boldsymbol{\Phi}(t)}{dt} &= (H_{eff} + H_{NL})\boldsymbol{\Phi}(t) + iW^T\boldsymbol{S_+}(t)\,, \label{Eq:tCMT-1}\\
  \boldsymbol{S_-}(t) &= -\boldsymbol{S_+}(t) + W\boldsymbol{\Phi}(t)\,, \label{Eq:tCMT-2}
\end{align}
where the components of the time-dependent vector $\boldsymbol{\Phi}(t)=(\phi_1,\phi_2,\cdots,\phi_N)e^{-i\omega t}$ describe the field amplitude at each mode $n=1,2,\cdots,N$ and we normalize $|\phi_n^{(\alpha)}|^2$ to be the $n-th$ modal energy density.
We assume that the system is excited by a monochromatic incident wave  $\boldsymbol{S_{+}}(t)= {\bf I} e^{-i\omega t}$, where ${\bf I}=(A_1,A_2)^T$, $\omega$ is the frequency of the incident wave, and $|A_{\alpha}|^2$ is the incoming power at the $\alpha$-th port.
Similarly $\boldsymbol{S_{-}}(t)= {\bf O} e^{-i\omega t}$ is the outgoing wave, where ${\bf O}=(O_{1},O_{2})^T$ and $O_{\alpha}$ is the field amplitude at port $\alpha=1,2$.
Substitution of these expressions in Eq.~(\ref{Eq:tCMT-2}) leads to Eqs.~(\ref{Eq:MCMT-1},\ref{Eq:MCMT-2}) of the main text.

We rewrite the steady-state coupled-mode-theory (CMT) equations that describe the scattering process as following
\begin{align}
  &(\omega - H_{eff} - H_{NL})\Phi^{(\alpha)} = iW^TI^{(\alpha)}\,, \label{Eq:MCMT-1-App} \\
  &\boldsymbol{O}^{(\alpha)} = CI^{(\alpha)} + W\Phi^{(\alpha)}\,, \label{Eq:MCMT-2-App}
\end{align}
where $\Phi^{(\alpha)}$ and $I^{(\alpha)}$ are the scattering vector field and the incident field vector, respectively.
The effective Hamiltonian $H_{eff} = H-  i\frac{W^TW}{2}$ describes the wave dynamics in the (linear) complex scattering domain when it is coupled to ports while $(H_{NL})_{nm}=f(|\phi_N^{(\alpha)}|^2)\delta_{nN}\delta_{nm}$ describes the non-linear interactions affecting the $N-$th resonant mode.
The system-ports coupling is described by the matrix $W$ with elements $W_{n,\alpha}=\delta_{n,\alpha}w_{\alpha}$ ($n=1,\cdots,N$).
By solving for $\Phi^{(\alpha)}$ from Eq.~(\ref{Eq:MCMT-1-App}) and substituting into Eq.~(\ref{Eq:MCMT-2-App}), we get
\begin{align} \label{SCMT-App}
  \boldsymbol{O} = \big[-\mathbf{1} + iW(\omega - H_{eff} - H_{NL})^{-1}W^T \big]\boldsymbol{I} = S\boldsymbol{I}
\end{align}
which allows us to obtain the $|\phi_N^{(\alpha)}|^2$-dependent scattering function $S$.
Similar to the case of graphs, ${\rm x}_{\alpha}$ is a solution of an algebraic equation that depends on $A_{\alpha}$, and therefore, $S=S(A_{\alpha})$.

For the general systems described by the coupled mode theory formulated in Eqs.~(\ref{Eq:MCMT-1-App},\ref{Eq:MCMT-2-App}) (also given in the main text), we can perform the calculation of the nonlinear field $|\phi_N^{(\alpha)}|$ and subsequently the transmission formula following the same steps similar to the ones that we have followed in graphs.
We can further reformulate Eqs.~(\ref{Eq:MCMT-1-App},\ref{Eq:MCMT-2-App}) as follow:
\begin{align}
  &(\tilde{H} + \tilde{H}_{NL} + i\tilde{W}^T\tilde{W})\Phi^{(\alpha)} = 2i\tilde{W}^T \tilde{I}^{(\alpha)}\,,\\
  &\tilde{\boldsymbol{O}}^{(\alpha)} = C\tilde{I}^{(\alpha)} + \tilde{W}\Phi^{(\alpha)}\,,
\end{align}
where $\tilde{H} = (\omega - H)/w^2$ (we take $w = w_1=w_2$ in our modeling), $\tilde{H}_{NL} = -H_{NL}/w^2$, $\tilde{W} = W/w$, $\tilde{I}^{(\alpha)} = I^{(\alpha)}/w$, $\tilde{\boldsymbol{O}}^{(\alpha)} =\boldsymbol{O}^{(\alpha)}/w$.

This reformulation, allows us to ``match'' the CMT scattering expressions to the ones derived in the case of graphs.
Following the same methodology with the graph-analysis, we first solve for the nonlinear field $|\phi_N^{(\alpha)}|$ based on Eq.~(\ref{Eq:MCMT-1-App}) by separating the wave amplitudes associated with the linear and nonlinear modes.
Consequently, we get a cubic equation for the nonlinear field intensity (contrast Eq.~(\ref{Eq:G_L-G_NL}) to Eq.~(\ref{Eq:Cardano-3}) applying for graphs).
Once the amplitude of the field at the nonlinear mode is evaluated, it can be substituted into Eq.~(\ref{Eq:MCMT-1-App}), in order to get the waves on each mode.
Following the same procedure as the one that we have used in graphs (from Eq.~(\ref{Eq:phi_N_alpha}) to Eq.~(\ref{Eq:Tmax-lossy-all-App})), we substitute the waves into Eq.~(\ref{Eq:MCMT-2-App}) and get the corresponding transmittance.
Then we can calculate the maximum transmission for both lossless and lossy cases.
For a lossless CMT, we get an expression of the maximum transmittance versus \SAF{} which is given by Eq.~(\ref{Eq:T-max}).
In other words, we conclude that also here the maximum transmission follows the theoretical bound as for a graph.

On the other hand, a CMT modeling that incorporates losses at one of the modes, that differ from the nonlinear one or/and the ones that are used
to attached the leads, result in a breaking of the transmission bound versus \SAF{}. This is demonstrated with pink circles in Fig.~\ref{fig5}b
where we have added losses $\lambda_3$ on mode $n=3$ for a CMT model of $N=4$ (see Fig.~\ref{fig5}a). Furthermore, we have selected one
CMT realization and evaluated the parametric evolution of the maximum transmission versus \SAF{} as the losses at mode $n=3$ are increased
from 0 to a large value (see blue circles in Fig.~\ref{fig5}(b)). At the two extreme cases (zero loss and high-loss values) the maximum transmission
follows the theoretical bound, while at intermediate loss values, this bound is violated.
As the losses increase from zero, the maximum transmission is initially decreases while the \SAF{} increases.
At  some critical value of the loss, the maximum transmission revert its behavior and starts increasing while the \SAF{} following an opposite
trend and decreases. Eventually, at high losses, the maximum transmission is bounded again by the results of Eq.~(\ref{Eq:T-max}).

The CMT modeling can be modified appropriately in order to describe a RMT.
Specifically, the Hamiltonian that describes the modes of the scattering system is drawn from a Gaussian Orthogonal Ensemble (GOE).
The RMT modeling is completed by enforcing two additional inputs.
The first one involves the values of the coupling elements $w_1,w_2$  such that the RMT modeling takes into account system-specific
direct processes occurring at graphs.
The latter are encoded in the energy (or ensemble) averaged $S-$matrix.
A direct comparison between the  RMT and the graph scattering matrix in the linear domain gives $w_\alpha = \sqrt{\frac{1}{\pi}\frac{1 - |\langle S_{\alpha,\alpha}\rangle|}{1 + |\langle S_{\alpha,\alpha}\rangle|}}$.
The second information that is needed is the appropriate RMT modeling of the nonlinear coefficients that define the nonlinearity strength.
Equivalently, we identify the incident field amplitudes for which the RMT and the graph model, lead to a statistically equivalent nonlinear term.
By comparing the scattering functions of the graph and the RMT (see Eq.~(\ref{Sgraph}) and Eq.~(\ref{SCMT-App}), respectively) we get
\begin{align} \label{Eq:NLEqCMTGraph_suppl}
  \frac{2 f_{RMT}(\langle|\phi_N^{RMT}|^2\rangle)}{w_1^2} = f_G(\langle|\phi_N^G|^2\rangle)\,,
\end{align}
where $w_1=w_2$ in our case.
Expressing $\phi_N^{RMT}, \phi_N^G$ in terms of $A_{\alpha}^{RMT}, A_{\alpha}^{G}$ allows us to establish an equivalence between
the incident fields of the RMT and graphs models that produce the same nonlinear effects. For Kerr nonlinearity case, we have
\begin{align} \label{Eq:NLEqCMTGraph-Kerr}
  \frac{2\chi_{RMT}\langle |\phi_N^{RMT}|^2\rangle}{w_1^2} = k\chi_G\langle|\phi_N^G|^2\rangle\,.
\end{align}
For saturable nonlinearity, we have
\begin{widetext}
\begin{align}
  \frac{2\left(z_0^{RMT} - z_1^{RMT}/(1+\chi_{RMT}\langle |\phi_N^{RMT}|^2\rangle)\right)}{w_1^2}
	 = k\left(z_0^{G} - z_1^{G}/(1+\chi_{G}\langle |\phi_N^{RMT}|^2\rangle)\right)\,.
	 \label{Eq:NLEqCMTGraph-Saturable}
\end{align}
\end{widetext}

\section{Adding Loss on vertices connected to leads and/or on the nonlinear vertex}
\label{Appendix:LossOnLeadVertex}

\begin{figure*}
  \centering \includegraphics[width=0.8\linewidth]{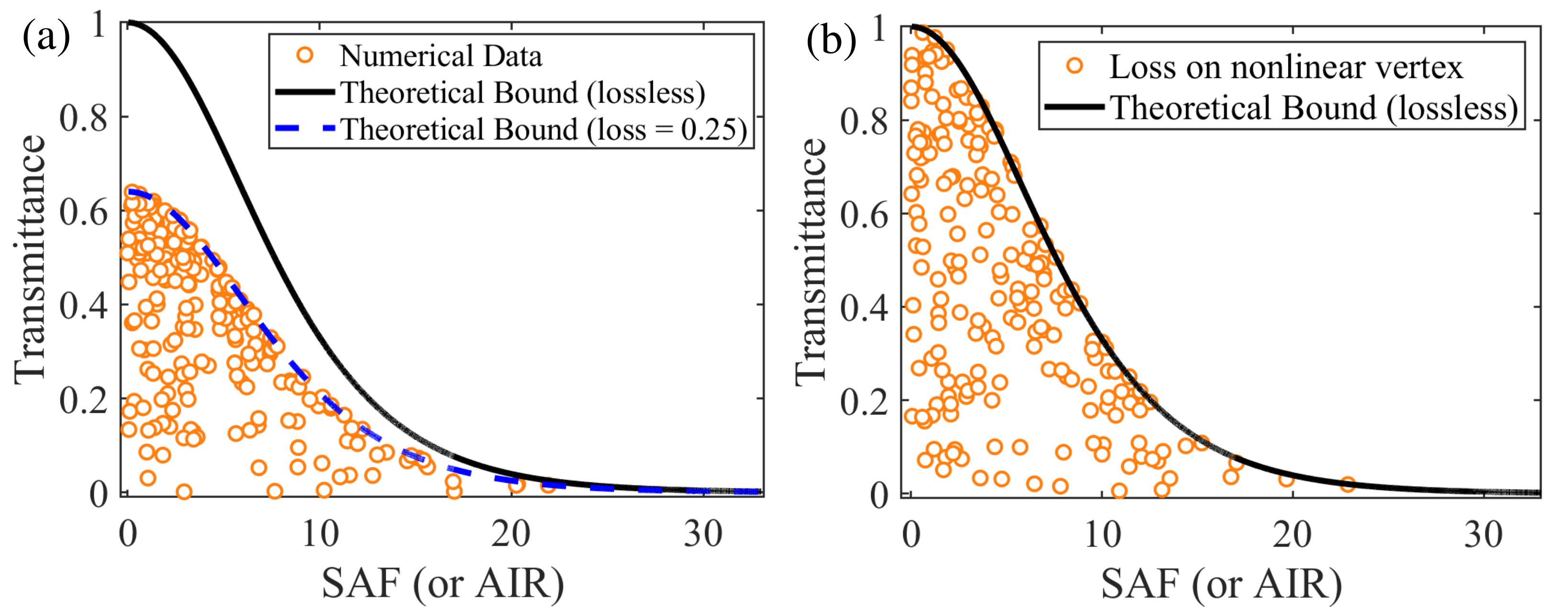}
  \caption{ \label{fig6}
  	{\bf Transmittance versus structural asymmetry factor (\SAF{}) --} (a) A graph with losses on vertices connected to leads and (b) a graph with
  	both linear and nonlinear (Kerr) losses on the nonlinear vertex.
  }
\end{figure*}

In the case that the losses $\lambda_{loss}$ are included in the vertices 1 and 2 that are connected with the leads 1 and 2 respectively, one needs
to modify the diagonal elements $M(1,1)$ and $M(2,2)$ of the matrix $M$ by adding the extra term $ik\lambda_{loss}$ on the left hand side of
the Eq.~(\ref{GMM}) (for simplicity we assume that the losses are the same in both vertices). For further theoretical processing we ``absorb'' these
extra terms to the graph-leads coupling matrix $iW^TW$. As a result, the left side of Eq.~(\ref{GMM}), takes the form $i(1+k\lambda_{loss})W^TW$.
We proceed by dividing both sides of Eq.~(\ref{GMM}) with the factor $(1+k\lambda_{loss})$.
Consequently the input amplitude appearing on the right hand side of Eq.~(\ref{GMM}) becomes $A_{\rm eff}=A/(1+k\lambda_{loss})$.
After performing the above manipulations, Eq.~(\ref{GMM}) is transformed to the following form:
\begin{widetext}
\begin{align}
  [(M + M_{NL})/(1+k\lambda_{loss}) + iW^TW]\Phi^{(\alpha)}
  = 2iW^TI^{(\alpha)}/(1+k\lambda_{loss}),
\end{align}
\end{widetext}
Consequently, the transmission formula Eq.~(\ref{genT}) will have an extra multiplicity factor $1/(1+k\lambda_{loss})^2$, reflecting the changes in
the effective input wave amplitude $A_{\rm eff}$. This rescaling of the input amplitude will affect also the whole transmission (and therefore
the maximum transmission) which now scales by the factor $1/(1+k\lambda_{loss})^2$ when compared to the lossless case. We have tested this theoretical
prediction via direct numerical simulations, see Fig.~\ref{fig6}a ($k=1$). Our detailed numerical analysis indicated that in cases, where these
losses $\lambda_1$ and $\lambda_2$ in vertices 1 and 2, respectively, are different from one another, the maximum transmission is bounded
by a similar factor as above with the substitution of $\lambda_{loss}=min\{\lambda_1,\lambda_2\}$. At the same time, we have checked via
detailed numerical simulations that in the case that the losses (linear or/and nonlinear) are introduced on the nonlinear vertex $N$ (here $N=4$)
the transmittance will be bounded by the expression given by Eq.~(\ref{Eq:T-max}), see Fig.~\ref{fig6}b.

\section{Proof of Identity 1 for lossless graphs}
\label{Appendix-proof1}

We will prove that in case of lossess graphs the following identity holds:
\begin{align} \label{Eq:equality2}
  |\Im{(b)}| = |c_1|^2 + |c_2|^2
\end{align}
We can express the imaginary part of $b$ defined in Eq.~(\ref{Definition-b}) as
\begin{widetext}
\begin{align} \label{Eq:Im_b}
	\Im(b) = -(M_{N1}, M_{N2},\cdots, M_{N,N-1})M_{N-1}^{-1}
	W_0(M_{N-1}+W_0M_{N-1}^{-1}W_0)^{-1}
	\begin{pmatrix}
		M_{1N} \\
		M_{2N} \\
		\vdots\\
		M_{N-1,N}
	\end{pmatrix}\,,
\end{align}
\end{widetext}
where the following matrix identity
\begin{widetext}
\begin{align} \label{matrixin}
	[M_{N-1}+iW_0]^{-1}=(M_{N-1} + W_0M_{N-1}^{-1}W_0)^{-1} - iM_{N-1}^{-1}
	W_0(M_{N-1}+W_0M_{N-1}^{-1}W_0)^{-1}
\end{align}
\end{widetext}
has been used.

At the same time one can express $|c_1|^2, |c_2|^2$ appearing in Eqs.~(\ref{Definition-c1}) as
\begin{widetext}
\begin{align}\label{Eq:identity-c1c2}
	&|c_1|^2 + |c_2|^2 = c_1c_1^{\dagger} + c_2c_2^{\dagger}=\\
		&(M_{N1}, M_{N2},\cdots, M_{N,N-1})
	\cdot(1+M_{N-1}^{-1}W_0M_{N-1}^{-1}W_0)(M_{N-1}+ W_0M_{N-1}^{-1}W_0)^{-1}
	W_0(M_{N-1}+W_0M_{N-1}^{-1}W_0)^{-1}
		\begin{pmatrix}
			M_{1N} \\
			M_{2N} \\
			\vdots\\
			M_{N-1,N}
		\end{pmatrix}\nonumber,
\end{align}
\end{widetext}
where in the derivation, we have utilized Eq.~(\ref{matrixin}) together with the equation $(M_{N-1}+W_0M_{N-1}^{-1}W_0)^{-1}
W_0M_{N-1}^{-1}=M_{N-1}^{-1}W_0(M_{N-1}+W_0M_{N-1}^{-1}W_0)^{-1}$ stemming from the fact that the transpose of a
symmetric matrix equals to itself ($M_{N-1}$ is symmetric and $W_0$ is diagonal with only first two elements nonzero).

At the same time $M_{N-1}^{-1}=(1+M_{N-1}^{-1}W_0M_{N-1}^{-1}W_0)(M_{N-1}+W_0M_{N-1}^{-1}W_0)^{-1}$ which can
be shown by multiplying from the right of this equality with $(M_{N-1}+W_0M_{N-1}^{-1}W_0)$. Substituting $M_{N-1}^{-1}$ to
Eq.~(\ref{Eq:Im_b}) allows us to show that $|\Im{(b)}| = |c_1|^2 + |c_2|^2$.\\

\section{Proof of identity 2 for lossless graphs}
\label{Appendix-proof}

We will prove that in case of lossess graphs the following identity holds:
\begin{align} \label{Eq:equality1}
  \Im(\frac{c_1c_2}{q}) = \Im(b)
\end{align}
We rewrite the above equality as follows:
\begin{align} \label{Eq:equality2_v2}
  \Im(c_1q^*c_2)=|q|^2\Im(b).
\end{align}
The left hand side of the above equation becomes
\begin{widetext}
\begin{equation} \label{Eq:c1qc2}
  c_1q^*c_2=(M_{N1}, M_{N2}, \cdots, M_{N, N-1})
  G
	\begin{pmatrix}
		1 \\
		0 \\
		\vdots\\
		0
	\end{pmatrix}
	(1, 0, \cdots, 0)G^*
	\begin{pmatrix}
		0 \\
		1 \\
		\vdots\\
		0
	\end{pmatrix}(0, 1, \cdots, 0)
	 G \begin{pmatrix}
		M_{1N} \\
		M_{2N} \\
		\vdots\\
		M_{N-1,N}
	\end{pmatrix}\,,
\end{equation}
\end{widetext}
where $G = [M_{N-1}+iW_0]^{-1}$ and we have used the definitions of $c_1,c_2,q$ appearing in Eqs.~(\ref{Definition-b},\ref{Definition-c1}).
The term $|q|^2$ in the above equation can be re-written as
\begin{align}\label{Eq:q2}
|q|^2 = \left|(1, 0, \cdots, 0)G
\begin{pmatrix}
0 \\
1 \\
\vdots\\
0
\end{pmatrix}\right|^2=\left|g_{12}\right|^2
\end{align}
where we have used the notation $g_{12} = (1, 0, \cdots, 0)G(0,1,\cdots,0)^T$.
Substituting Eqs.~(\ref{Eq:c1qc2}), (\ref{Eq:Im_b}), (\ref{Eq:q2}) into Eq.~(\ref{Eq:equality2_v2}) allows us to re-write the latter as following
\begin{align}
  g_{12}^* G I_0 G=- |g_{12}|^2 ZW_0 X\,,
\end{align}
where we have denoted $I_0 =(1, 0, \cdots, 0)^T(0, 1, 0, \cdots, 0), X = [M_{N-1} + W_0ZW_0]^{-1}$ and $Z = M_{N-1}^{-1}$.
This expression can further collapse to the following form
\begin{widetext}
\begin{align}
  \Im(g_{12}^*)I_0 - \Im(g_{12}^*)W_0ZI_0ZW_0 - \Re(g_{12}^*)I_0Z W_0 - \Re(g_{12}^*W_0ZI_0)
  =-|g_{12}|^2(W_0 + W_0ZW_0ZW_0)\,.
\end{align}
\end{widetext}
which can be explicitly written in matrix form as:
\begin{widetext}
\begin{align} \label{Eq:2by2matrix}
\begin{pmatrix}
0 & g_{12}^{*i}  \\
0 & 0
\end{pmatrix}-g_{12}^{*i}
\begin{pmatrix}
z_{11}z_{21} & z_{11}z_{22}  \\
z_{21}z_{21} & z_{21}z_{22}
\end{pmatrix} - g_{12}^{*r}
\begin{pmatrix}
z_{21} & z_{22}  \\
0 & 0
\end{pmatrix}- g_{12}^{*r}
\begin{pmatrix}
0& z_{11}  \\
0 & z_{21}
\end{pmatrix}
=-|g_{12}|^2
\begin{pmatrix}
1& 0  \\
0 & 1
\end{pmatrix}-|g_{12}|^2
\begin{pmatrix}
z_{11}^2+z_{12}^2& z_{11}z_{12}+z_{12}z_{22}  \\
z_{12}z_{11}+z_{12}z_{22} & z_{12}^2+z_{22}^2
\end{pmatrix}\,,
\end{align}
\end{widetext}
where $z_{mn}$ are the $(m,n)$ matrix element of the matrix $Z$.

Finally, using the relation $G(1+W_0M_{N-1}^{-1}W_0M_{N-1}^{-1})=M_{N-1}^{-1} - iM_{N-1}^{-1}W_0M_{N-1}^{-1}$ we
can extract the connection between the matrix element $g_{12}$ and the elements of $Z$ as
\begin{widetext}
\begin{align}
g_{12}^r \equiv \Re(g_{12}) &= \frac{z_{12}(z_{11}z_{22} - z_{12}^2-1)}{z_{12}^2(z_{11}+z_{22})^2 -
(z_{11}^2+z_{12}^2+1)(z_{12}^2 + z_{22}^2 +1)} \label{Eq:t12r}\\
g_{12}^i \equiv \Im(g_{12}) &= \frac{z_{12}(z_{11}+z_{22})}{z_{12}^2(z_{11}+z_{22})^2 - (z_{11}^2+
z_{12}^2+1)(z_{12}^2 + z_{22}^2 +1)} \label{Eq:t12i} \\
|g_{12}|^2 &= \frac{z_{12}^2[(z_{11}z_{22} - z_{12}^2-1)^2+(z_{11}+z_{22})^2]}{[z_{12}^2(z_{11}+
z_{22})^2 - (z_{11}^2+z_{12}^2+1)(z_{12}^2 + z_{22}^2 +1)]^2}\,.
\label{Eq:t12sq}
\end{align}
\end{widetext}
Substituting the formulas (\ref{Eq:t12r}), (\ref{Eq:t12i}), and (\ref{Eq:t12sq}) back to Eq.~(\ref{Eq:2by2matrix}), we can  prove its validity
and therefore the validity of Eq.~(\ref{Eq:equality1}).

\section{Enhanced AIR in lossy systems due to resonant mode overlapping and/or $\Im{\left(\Lambda\right)}<1$}\label{TRSV}

\begin{figure*}
  \centering \includegraphics[width=\linewidth]{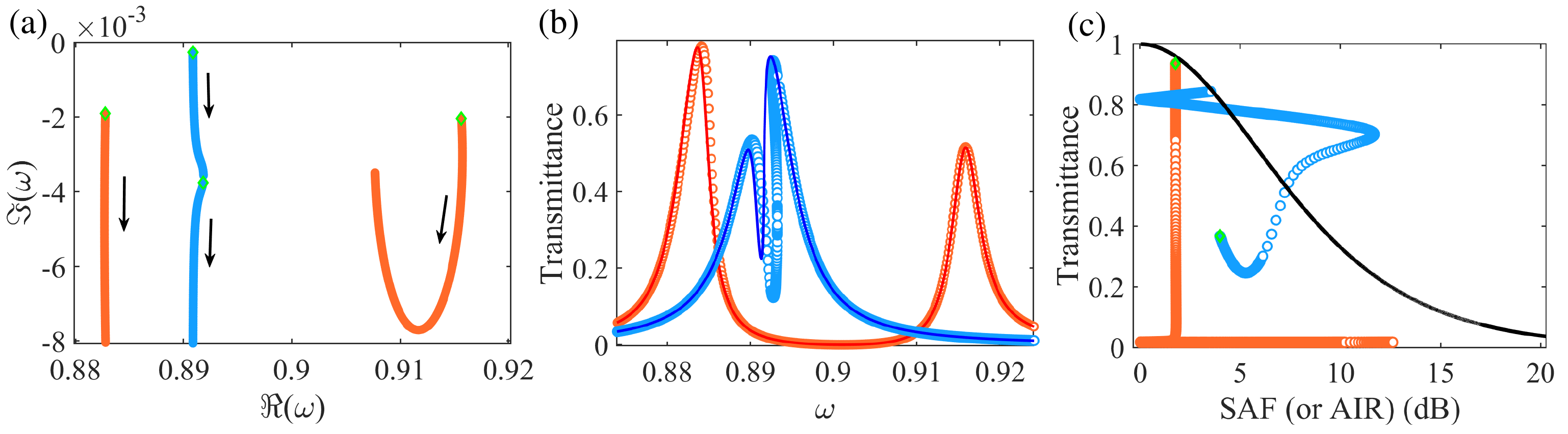}
  \caption{ \label{fig7}
{\bf Resonant mode overlapping effect in AT for an RMT model -} A RMT model consisting of 6 modes with a variable loss added on mode $n=3$
and a Kerr-nonlinearity added on mode $N=6$. The coupling constant of the system with the leads is $w_{\alpha=1,2}=0.1$. (a) Parametric evolution
of linear resonance modes in the complex frequency plane as the variable loss increases from zero (green diamond) to some value. The black arrows
indicate the direction of the resonant motion as the loss is increased. The orange (blue) circles correspond to two distinct RMT models whose only
difference is one coupling element $H_{4,5}$. In the former case $H_{4,5}$ is such that the resonances do not overlap while in the latter
case they form a quasi-degenerate pair.  (b) The nonlinear transmission spectrum for these two RMT models. The imaginary part (loss strength) of
the $n=3$-resonant mode takes the value $0.0006$. Colored symbols correspond to the associated RMT model that has been used in subfigure (a).
The RMT model that supports the quasi-degenerate pair of resonant modes (blue circles) shows a larger differences between the left (blue circles)
$T_1$ and right (solid blue line) $T_2$ transmittances than the corresponding ones associated with the RMT model where the modes are isolated. (c)
The transmission versus SAF (solid black line) for the two cases discussed previously for a fixed value of the frequency of the incident wave ($\omega
=0.884$ for the isolated resonance RMT model and $\omega=0.892$ for the quasi-degenerate resonance model). The loss changes from zero (green
diamonds) to the same maximum value as the one used in subfugure Fig. \ref{fig7}a. The system that supports quasi-degerenate resonances
break the lossless bound Eq. (\ref{Eq:T-max}) at certain loss values. In both (b,c) the input amplitude is $A=0.2$, and the nonlinear coefficient
is $\chi = 0.01$.
  }
\end{figure*}

An important consequence of the addition of losses is the broadening of the resonance line-width. It turns out that the lossless bound Eq. (\ref{Eq:T-max})
is violated whenever two resonances interact with one-another like in the case of non-reciprocal transport induced via magnetic field in the presence
of losses. The phenomenon is more profound when these resonances create a quasi-degenerate pair. This scenario is better illustrated in Fig. \ref{fig7}
where we have analyzed the resonant mode behavior and transmittance of a non-linear RMT model. The system consists of six resonance modes
which were coupled with one-another via random couplings. A Kerr-nonlinearity was assumed to act in resonant mode $N=6$. Variable losses
have been introduced in resonant mode $n=3$. We have realized two replicas of this system that differ from one-another by one coupling element.
The choice of this element is such that in one case (blue circles) the two resonances form a quasi-degenerate pair as opposed to the other case (orange
circles) where they are well separated. As the losses are increased the quasi-degenerate pair of resonances overlap strongly and interact with one another
via the nonlinear term. This nonlinear interaction enforces strong interference effects which amplify the asymmetric transport (see Fig. \ref{fig7}b)
and induce a violation of the lossless bound for maximum transmittance Eq. (\ref{Eq:T-max}). The latter is clearly seen in Fig. \ref{fig7}c where we
plot the transmittance for each of these cases at a fixed frequency and varying losses.

\begin{figure*}
  \centering \includegraphics[width=1\linewidth]{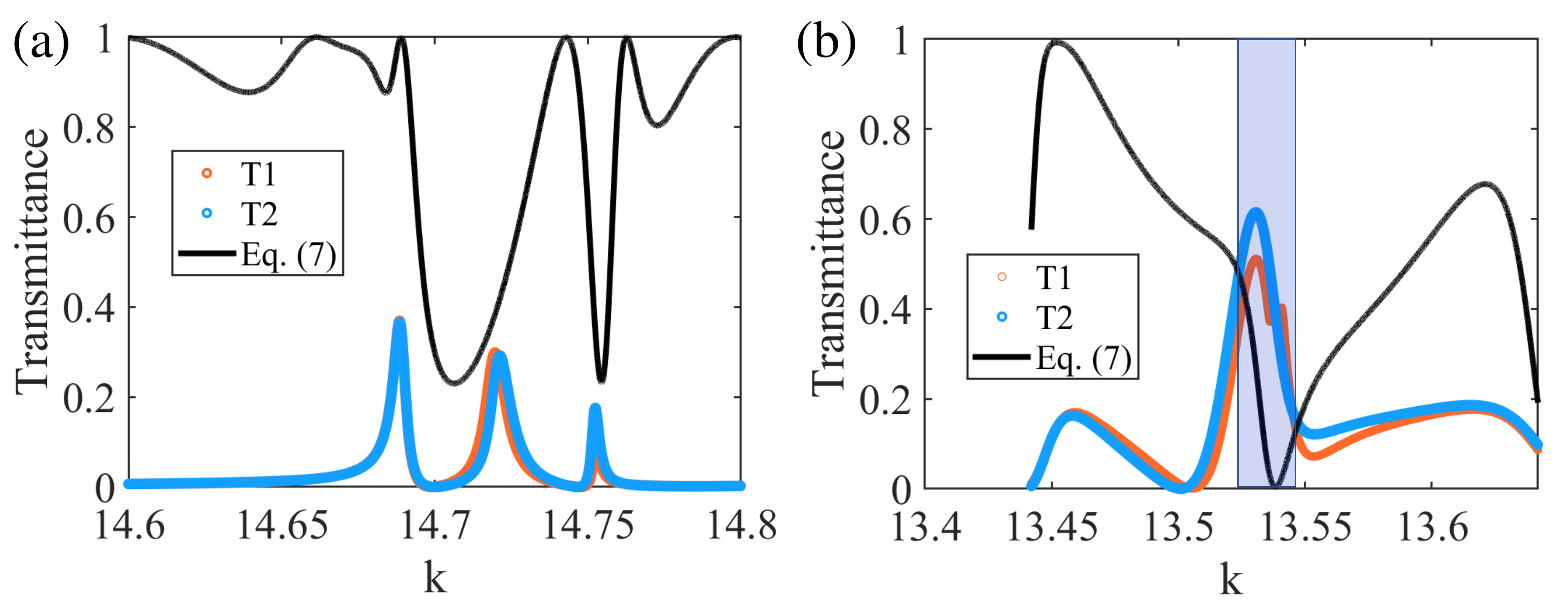}
  \caption{ \label{fig8}
{\bf Resonance mode overlapping effect in AT for a graph -} Transmission spectrum of a tetrahedron graph used in the insets of Fig.~\ref{fig2}(b).
The red (blue) circles indicate the left (right) transmittances of a nonlinear lossy graph with a dielectric constant $\Im{\lambda}_3=0.15$ at
vertex $n=3$, and a Kerr nonlinear coefficient $\chi = 1$ at the vertex $N=4$. (a) The graph supports isolated resonances and both transmittances
(input amplitude $A=20$) are below the maximum bound of Eq. (\ref{Eq:T-max}), i.e., $T_{1,2}<T_{\rm max}$. (b) The same as in (a), but now
the graph supports overlapping resonances (input amplitude $A=10$). In this case, $T_{1,2}>T_{\rm max}$ (see blue highlighted area). The black
curves in both subfigures indicate the corresponding maximum transmittance bound given by Eq.~(\ref{Eq:T-max}).
  }
\end{figure*}

In Fig. \ref{fig8} we report a similar scenario for the graph configuration that we have used in the insets of Fig. \ref{fig2}b. It consists of four
vertices with a Kerr-nonlinearity at the $N=4$ vertex and a lossy dielectric constant at $n=3$ i.e. $\Im({\lambda}_3)=0.15$. In order to make
clearer our point we have also introduced additional real-value dielectric constants $\lambda_{1,2}=0.5$ at the vertices $n=1,2$ where the TLs
are attached. These ``electrical potential barriers'' enforce the formation of well isolated resonances, even in the presence of losses. This scenario
is depicted in Fig. \ref{fig7}a where the left $T_1$ and right transmittance  $T_2$ of such lossy graph is shown with red and blue circles,
respectively. At the same figure, we show the corresponding maximum transmittance $T_{\rm max}$ (black line) given by Eq. (\ref{Eq:T-max}).
Both $T_1$ and $T_2$ are below $T_{\rm max}$. Instead, in Fig. \ref{fig7}b, we have eliminated the electrical barriers $\lambda_{1,2}=0$.
In this case, the resonance modes overlap, leading to transmittances that violate the upper bound given by Eq. (\ref{Eq:T-max}) (see blue
highlight domain).

\begin{figure*}
  \centering \includegraphics[width=0.5\linewidth]{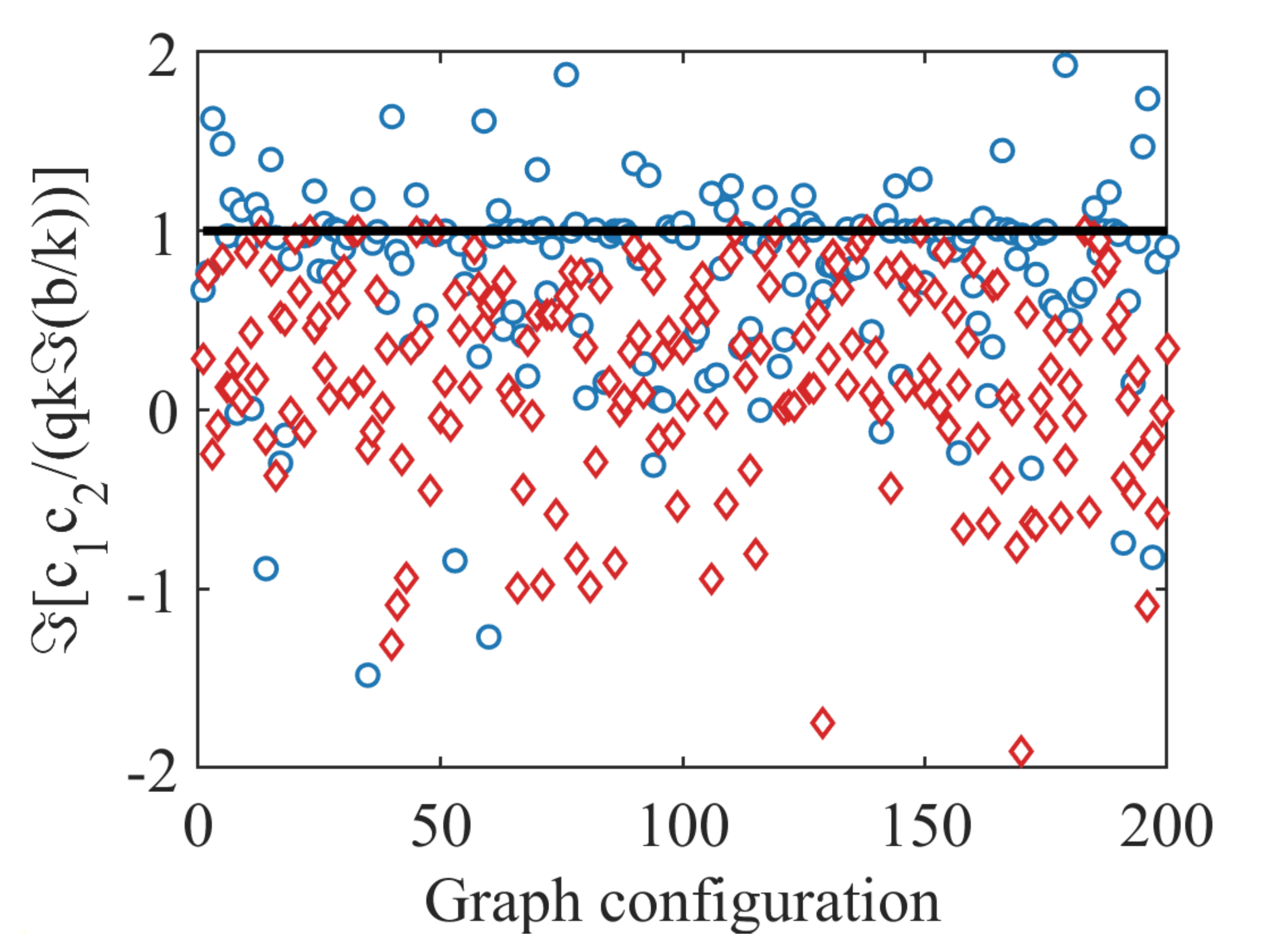}
  \caption{ \label{fig9}
{\bf  Implications of $\Im{\left(\Lambda\right)}<1$-} Monte-Carlo simulations using a lossy tetrahedron graph for various wavevector
$k$-values, bond-length configurations, loss-strengths etc. In all cases that the inequality $\Im{\left(\Lambda\right)}<1$ is satisfied
(red diamonds), the transmittance exceeds the value given by Eq.~(\ref{Eq:T-max}). The blue circles are all possible graph
configurations for which the bound of Eq.~(\ref{Eq:T-max}) is not violated. In such cases we do not expect enhanced AIR (for fixed
$T_{\rm max}$) or enhanced transmission asymmetry bound (for fixed AIR) than the one given by Eq. (\ref{Eq:T-max}). The black
horizontal line indicates the equality $\Im{\left(\Lambda\right)}=1$.
  }
\end{figure*}

Finally, we present numerical results on the consequences of the inequality $\Im{\left(\Lambda\right)}<1$. From Eq.
(\ref{Eq:Tmax-lossy}) we speculate that if $\Im{\left(\Lambda\right)}<1$, the lossy graph configurations might violate the lossless
bound Eq. (\ref{Eq:T-max}), leading to enhanced AIR (for fixed $T_{\rm max}$) or enhanced transmission asymmetry bound (for
fixed AIR) than the one given by Eq. (\ref{Eq:T-max}). In Fig. \ref{fig9} we present some Monte-Carlo simulations with a tetrahedron
graph (for various $k$-values, length configurations etc), which confirmed that the above inequality is a necessary but not sufficient
condition for violating the lossless limit of Eq. (\ref{Eq:T-max}).

\end{document}